\documentclass[aps, prb, reprint, preprintnumbers,superscriptaddress,nofootinbib,longbibliography]{revtex4-2}

\usepackage[utf8]{inputenc}
\usepackage[english]{babel}

\usepackage{graphicx}

\usepackage{amsfonts,amssymb,amsmath}
\usepackage{mathtools}

\usepackage{bm}
\usepackage{times}
\usepackage{units}
\usepackage{dsfont}
\usepackage{mathrsfs}

\usepackage{array}
\usepackage{tabularx}
\usepackage{multirow}

\usepackage{hyperref}
\hypersetup{colorlinks=true,linktoc=all,linkcolor=blue,breaklinks=true,citecolor=blue,urlcolor=blue}

\usepackage[normalem]{ulem}
\usepackage{xspace}
\usepackage[dvipsnames]{xcolor}

\addto\captionsenglish{}

\AtBeginDocument{%
    \newwrite\bibnotes
    \def\bibnotesext{Notes.bib}
    \immediate\openout\bibnotes=\jobname\bibnotesext
    \immediate\write\bibnotes{@CONTROL{REVTEX42Control}}
    \immediate\write\bibnotes{@CONTROL{%
            apsrev42Control,author="08",editor="1",pages="1",title="0",year="1"}}
    \if@filesw
    \immediate\write\@auxout{\string\citation{apsrev42Control}}%
    \fi
}%

\newcommand*{\e}{\text{e}}
\newcommand*{\ii}{\text{in}}
\renewcommand*{\L}{\text{L}}
\newcommand*{\R}{\text{R}}
\newcommand*{\kB}{k_\text{B}}
\newcommand*{\M}{\text{M}}
\newcommand*{\Tr}{\text{Tr}}
\newcommand*{\eq}{\text{eq}}
\newcommand*{\Deq}{D}
\newcommand*{\Feq}{F_\eq}
\newcommand*{\Fneq}{F}
\newcommand*{\Peq}{P^\eq}
\newcommand*{\bPeq}{\bar{P}^\eq}
\newcommand{\rhoeq}{\rho^\eq}
\newcommand{\brhoeq}{\bar{\rho}^\eq}
\newcommand*{\gamp}{\gamma_p}
\newcommand*{\gamc}{\gamma_c}
\newcommand*{\gams}{\gamma_s}
\newcommand{\one}{\mathds{1}}
\newcommand*{\Neq}{N_\eq}
\newcommand*{\bNeq}{\bar{N}_\eq}
\newcommand*{\bdNeq}{\delta\bar{N}_\eq^2}

\DeclarePairedDelimiter{\abs}{\lvert}{\rvert}

\DeclarePairedDelimiter{\ket}{\lvert}{\rangle}
\DeclarePairedDelimiterX{\proj}[1]{\lvert}{\rvert}{#1\,\delimsize\rangle\mathopen{}\delimsize\langle\,\mathopen{}#1}

\DeclarePairedDelimiter{\Bra}{{\bm{(}}}{{\bm{\rvert}}}
\DeclarePairedDelimiter{\Ket}{{\bm{\lvert}}}{{\bm{)}}}
\DeclarePairedDelimiterX{\Braket}[2]{{\bm{(}}}{{\bm{)}}}{#1\delimsize\bm{\vert}\mathopen{}#2}
\DeclarePairedDelimiterX{\Ketbra}[2]{{\bm{|}}}{{\bm{|}}}{#1\delimsize\bm{)}\mathopen{}\delimsize\bm{(}\mathopen{}#2}
\DeclarePairedDelimiterX{\Mel}[3]{{\bm{(}}}{{\bm{)}}}{#1\,\delimsize\bm{\vert}\mathopen{}#2\delimsize\bm{\vert}\,\mathopen{}#3}

\newcommand{\subfigref}[2]{\ref{#1}\hyperref[#1]{(#2)}}

\begin{document}

\title{Role of electron-electron interaction in the Mpemba effect in quantum dots}

\author{Juliane Graf}
\affiliation{Institute for Theoretical Physics, University of Regensburg, D-93053 Regensburg, Germany}
\affiliation{Department of Microtechnology and Nanoscience (MC2), Chalmers University of Technology, S-412 96 G\"oteborg, Sweden}
\affiliation{Department of Engineering and Physics, Karlstad University, Karlstad, Sweden}

\author{Janine Splettstoesser}
\affiliation{Department of Microtechnology and Nanoscience (MC2), Chalmers University of Technology, S-412 96 G\"oteborg, Sweden}

\author{Juliette Monsel}
\affiliation{Department of Microtechnology and Nanoscience (MC2), Chalmers University of Technology, S-412 96 G\"oteborg, Sweden}
\date{\today}

\begin{abstract}
    The Mpemba effect has initially been noticed in macroscopic systems---namely that hot water can freeze faster than cold water---but recently its extension to open quantum systems has attracted significant attention. This phenomenon can be explained in the context of nonequilibrium thermodynamics of Markovian systems, relying on the amplitudes of different decay modes of the system dynamics. Here, we study the Mpemba effect in a single-level quantum dot coupled to a thermal bath, highlighting the role of the sign and magnitude of the electron-electron interaction in the occurrence of the Mpemba effect. We gain physical insights into the decay modes from a dissipative symmetry of this system called fermionic duality. Based on this analysis of the relaxation to equilibrium of the dot, we derive criteria for the occurrence of the Mpemba effect using two thermodynamically relevant measures of the distance to equilibrium, the nonequilibrium free energy and the dot energy. We furthermore compare this effect to a possible exponential speedup of the relaxation. Finally, we propose experimentally relevant schemes for the state preparation and explore different ways of observing the Mpemba effect in quantum dots in experiments.
\end{abstract}

\maketitle

\section{Introduction}

The classical Mpemba effect refers to the counterintuitive phenomenon that hot water can freeze faster than the same amount of colder water. This effect has been observed throughout history for water freezing~\cite{Aristotle,Descartes,Mpemba1969May, Kell1969May} but more recently also in other systems~\cite{Ahn2016Jun, Lasanta2017Oct, Hu2018Oct}. However, the complexity of a liquid like water undergoing a phase transition makes it difficult to understand the mechanisms at play, such that the phenomenon remains debated~\cite{Bechhoefer2021Aug}. The Mpemba effect is easier to comprehend and verify by recasting it in the context of classical statistical physics of Markovian systems, where it becomes an anomalous relaxation process and where it can be explained by analyzing the decay modes of the evolution of the system: A smaller overlap with the slowest decay mode leads to a faster decay even if the initial temperature is hotter~\cite{Lu2017May, Klich2019Jun}. This was for instance observed in colloidal systems~\cite{Kumar2020Aug}.
The Mpemba effect has been further generalized, beyond systems with well-defined temperatures, to include any situation where a system relaxes faster toward a steady state from an arbitrary initial state far away from this steady state than it relaxes from another, closer, initial state. This definition can be straightforwardly applied to quantum open systems~\cite{Chatterjee2023Aug, Murciano2024Jan, Moroder2024Apr, Strachan2024Apra, Joshi2024Jul, Rylands2024Jul, Chatterjee2024Aug, Yamashika2024Auga, Nava2024Sep, Liu2024Oct, Liu2024Oct}, where a variety of systems and distance measures between two initial states has been considered~\cite{Biswas2023Aug}. The increased level of abstraction also allowed linking the Mpemba effect to other interesting concepts such as symmetry restoration~\cite{Joshi2024Jul, Rylands2024Jul, Liu2024Oct}, entanglement asymmetry~\cite{Murciano2024Jan, Yamashika2024Auga} , or exceptional points~\cite{Chatterjee2024Aug}. At the same time, physical intuition is limited by this increased abstraction and the connection to thermodynamics becomes less obvious. Also, for Markovian quantum systems, the Mpemba effect has been observed so far only in ion traps \cite{Zhang2024Jan, AharonyShapira2024Jul}.

An alternative promising platform to investigate the Mpemba effect is single-level quantum dots. They are simple systems but exhibit rich physics due to electron-electron interaction and level degeneracies, providing different decay modes. Furthermore, counting experiments allow revealing the time-evolution of the full quantum-dot state \cite{Guttinger2008Nov,Fricke2011Dec,Volk2013Apr,Hartman2018Nov,Garreis2023Jan}, thereby giving access to anomalous relaxation effects.
The Mpemba effect in a single-level quantum dot has been theoretically analyzed~\cite{Chatterjee2023Aug, Wang2024Sep} and identified both in an effective time-dependent quantum-dot temperature and in the separate density matrix elements~\cite{Chatterjee2023Aug}. However, the physical meaning of the relaxation rates and modes has neither been elucidated nor exploited in this analysis.

In the present manuscript, we study the Mpemba effect in a single-level quantum dot but extend our analysis to arbitrary local electron-electron interaction reaching from the strongly repulsive case of Ref.~\cite{Chatterjee2023Aug} to non-interacting dots and quantum dots with an effective attractive electron-electron interaction. We thereby elucidate the role of interaction for the Mpemba effect, both concerning the impact of interactions on decay rates and concerning the type of observables in which the Mpemba effect is visible. We find that a strong attractive interaction, which was emulated experimentally in Refs.~\cite{Cheng2015May,Cheng2016Dec,Hamo2016Jul,Prawiroatmodjo2017Aug} and can be mediated by phonons \cite{Koch2007May}, is particularly favorable to observing the Mpemba effect due to the large difference between the two relevant decay rates.

Using a so-called fermionic duality relation~\cite{Schulenborg2016Feb,Schulenborg2017Dec,Monsel2022Jul}, which holds for the time-evolution operator of the dissipative quantum-dot dynamics, we identify the charge and parity mode as the relevant modes that need or need not to be excited in order for the (strong) Mpemba effect to occur.
We analyze different measures~\cite{Biswas2023Aug} which reveal the dynamics of these modes and where the Mpemba effect can hence be visible. On one hand, to keep the connection with thermodynamics, we study the case of a relaxation toward an equilibrium state measuring the distance to equilibrium based on the nonequilibrium free energy~\cite{Moroder2024Apr}, which is equivalent to analyzing the relative entropy. The nonequilibrium free energy or the relative entropy are relevant measures in the context of resource theory for thermodynamics~\cite{Brandao2013Dec}. On the other hand, we show that for interacting systems, the Mpemba effect can be revealed from the dynamics of the internal quantum-dot energy, which is of experimental relevance, e.g. for the dynamics of quantum batteries~\cite{Campaioli2024Jul} or quantum heat engines~\cite{Kosloff2014Apr, Benenti2017Jun, Cangemi2024Oct, Lin2022Jan}.
We finally provide experimentally relevant protocols to prepare pairs of initial states exhibiting the Mpemba effect via a rapid gate switch or a change of the bath temperature.

The paper is organized as follows. We introduce the general theoretical framework in Sec.~\ref{sec:framework}, where we define the Mpemba effect (Sec.~\ref{sec:def Mpemba}), present the master equation describing the dynamics of a single-level quantum dot in contact with a thermal bath (Sec.~\ref{sec:model dot}), analyze the decay modes involved in the relaxation (Sec.~\ref{sec:relax}) and apply those insights to the observables relevant to the Mpemba effect (Sec.~\ref{sec:Mpemba_relax}).
Then, in Sec.~\ref{sec:Mpemba}, we investigate in detail the Mpemba effect in the quantum dot, based first on the nonequilibrium free energy and second on the internal energy. Finally, we propose protocols to prepare initial states showing a Mpemba effect in Sec.~\ref{sec:init state}.

\section{Quantum-dot dynamics and observables}\label{sec:framework}

\subsection{Mpemba effect in open quantum systems}\label{sec:def Mpemba}

Roughly speaking, the Mpemba effect occurs if a ``hotter'' initial state decays faster to equilibrium (or to the steady state) than a ``colder'' initial state. Concretely this means that a crossing occurs in the decay of an observable which quantifies how ``hot'' or ``cold'' a state is. This effect can be combined with (but has to be distinguished from) an exponential speedup of the relaxation of the ``hotter'' system to the steady state. Exponential speedup means that the decay of one state is exponentially faster than the decay of another state.  If both effects occur at the same time, one talks about a \textit{strong} Mpemba effect.
When adapting the analysis of the Mpemba effect to open quantum systems, the concept of ``hotter'' and ``colder'' initial states must be generalized in a way that is meaningful for small systems that are not in a thermal state.
In this paper, we follow two different paths to reveal the occurrence of the Mpemba effect in quantum dots.

First, we analyze the nonequilibrium free energy or the closely related relative entropy, which both are good measures of how far a system is from its thermal state.
When analyzing  the nonequilibrium free-energy difference $\Delta \Fneq$, following Ref.~\cite{Moroder2024Apr}, we use
\begin{equation}\label{def DF}
    \Delta\Fneq[\rho(t)]= \Fneq[\rho(t)] - \Feq = \kB T \Deq[\rho(t)] ,
\end{equation}
to quantify the distance of the system in state $\rho(t)$ at time $t$ to the equilibrium $\rhoeq$.
Here, we have introduced the equilibrium free energy $\Feq = -\kB T \log Z$, the Boltzmann constant $\kB$, the temperature $T$ to which the system relaxes, and the system's partition function $Z$, as well as the relative entropy with respect to the equilibrium state,
\begin{equation}\label{def_Deq}
    \Deq(\rho) = \Tr[\rho(\log\rho - \log\rhoeq)].
\end{equation}
To avoid involving the final temperature, the relative entropy alone can be taken as an equivalent measure for how far the state is from equilibrium. The Mpemba effect occurs for a pair of initial states $\varrho(0)$ and $\varrho'(0)$ if
\begin{subequations}\label{Mpemba_general}
\begin{align}
    \Deq[\varrho'(0)] &> \Deq[\varrho(0)],\label{Mpemba_1}\\
    \exists\, t_\M > 0:\; \forall\,t>t_\M:\;\Deq[\varrho'(t)] &< \Deq[\varrho(t)].\label{Mpemba_2}
\end{align}
\end{subequations}
Condition \eqref{Mpemba_1} expresses that $\varrho'(0)$ is ``hotter'', namely farther from the equilibrium, than $\varrho(0)$, while condition  \eqref{Mpemba_2} indicates that $\varrho'$ relaxes faster than $\varrho$ such that it is always closer to the equilibrium state after time $t_\M$. In the following, we will use calligraphic letters, like $\varrho$ and $\varrho'$ in Eq.~\eqref{Mpemba_general}, to refer to specific states for which we investigate the Mpemba effect, while standard letters are used to refer to generic states of the system, like $\rho$ in Eq.~\eqref{def_Deq}.
The choice of the nonequilibrium free-energy difference or the relative entropy as measures for how far the system is from the equilibrium is motivated by the fact that both quantities are zero only when the system is in equilibrium, and that both functions are monotonically decreasing with time during the relaxation process.

Even though the relative entropy (or the nonequilibrium free energy difference) has appealing properties, it is not the only way to define how far a state is from equilibrium for investigating the Mpemba effect. In the original Mpemba effect~\cite{Mpemba1969May}, one could simply measure the time evolution of the temperature of the studied, \textit{macroscopic} system, which might motivate to define an effective temperature of the dot~\cite{Chatterjee2023Aug}. Here, we instead study the decay of the internal energy of the quantum dot as a second means to diagnose the Mpemba effect. The internal energy is a \textit{measurable} observable which is of direct practical relevance in thermodynamic devices such as engines~\cite{Benenti2017Jun} or batteries~\cite{Campaioli2024Jul}.
Concretely, the energy stored in the quantum dot is evaluated from the time-dependent expectation value of the system Hamiltonian, which will be introduced in the following section. Methods to experimentally detect this decay are for example time- and energy-resolved current measurements~\cite{Fletcher2019Nov,Schulenborg2024Mar}.

The Mpemba effect occurring in these two observables of choice can be understood by investigating the eigenmodes and eigenvalues of the time-evolution operator describing the dynamics of the system. In the following, we will therefore set up the quantum-dot model and analyze its dynamics relaxing from a nonequilibrium state to equilibrium.

\subsection{Single-level quantum dot}\label{sec:model dot}

\begin{figure}[t]
    \includegraphics[width=\linewidth]{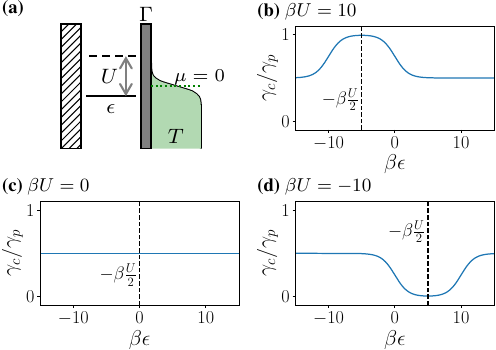}\vspace{-0.2cm}
    \caption{\textbf{(a)} Single-level quantum dot with energy level $\epsilon$ and electron-electron interaction $U$ (we here sketch the repulsive case, $U > 0$, as an example), coupled with tunnel rate $\Gamma$ to an electronic reservoir at the temperature $T$ and the electrochemical potential $\mu$, which is used as the energy reference, i.e., $\mu \equiv 0$. \textbf{(b), (c), (d)}  Ratio between charge rate and parity rate as a function of $\epsilon$ for \textbf{(b)} strong repulsive interaction, \textbf{(c)} no interaction and \textbf{(d)} strong attractive interaction. The dashed vertical line indicates the electron-hole symmetric point $\epsilon = -U/2$.}
    \label{fig:rates}
\end{figure}

We are interested in the dynamics of a single-level quantum dot coupled to a single electronic reservoir, see Fig.~\subfigref{fig:rates}{a}. Such a system can be described  by the Anderson impurity model \cite{Anderson1961Oct}, where the Hamiltonian is given by $H_\text{tot} = H +  H_\text{res} + H_\text{tun}$. Here, we have defined
\begin{subequations}
    \begin{align}
        {H}&=\sum_{\sigma}\epsilon{N}_\sigma+U{N}_\uparrow{N}_\downarrow,\label{H_dot}\\
        {H}_\text{res}&=\sum_{k,\sigma}\varepsilon_{k\sigma}^{} {c}_{ k\sigma}^\dagger{c}_{ k\sigma}^{}\label{H_lead},\\
        {H}_\text{tun}&=\sum_{ k,\sigma}V_{k\sigma}^{}{c}_{ k \sigma}^\dagger{d}_\sigma^{}+\text{h.c.}\label{H_tun},
    \end{align}
\end{subequations}
with ${d}_\sigma$ the dot annihilation operator for spin $\sigma=\;\uparrow,\downarrow$ and $N_\sigma = {d}_\sigma^\dagger {d}_\sigma^{}$. The Hamiltonian of the isolated quantum dot, $H$, contains not only the single-level energy $\epsilon$, but also an interaction energy $U$, which we will here consider to be positive (repulsive, Coulomb interaction), zero (noninteracting), or negative (effective attractive interaction, such as realized in Refs.~\cite{Cheng2015May, Cheng2016Dec,Hamo2016Jul,Prawiroatmodjo2017Aug}). Electrons in the reservoir are described by the noninteracting Hamiltonian ${H}_\text{res}$.
From the tunneling Hamiltonian, ${H}_\text{tun}$, we furthermore define the tunneling rates
\begin{equation}
    \Gamma_{\sigma}(E) = 2\pi \sum_k \delta(E - \varepsilon_{ k \sigma})\abs{V_{ k \sigma}}^2.
\end{equation}
In the spin-degenerate, wideband limit, the tunneling rates are constant and equal, $\Gamma_{\sigma}(E) = \Gamma$.

In this work, we consider the experimentally relevant weak-coupling limit $\hbar\Gamma \ll \kB T$ and $\Gamma \tau_c \ll 1$, where $\tau_c$ is the correlation time in the reservoir. In this regime, the exact magnitude of the tunneling rate $\Gamma$ never plays a role as long as time-dependent quantities are plotted as functions of $\Gamma t$. Furthermore, the evolution of the
coherences decouples from the one of the populations in the energy eigenbasis and we can therefore disregard the off-diagonal elements of the density matrix. We can then describe the dot state by a master equation for the evolution of the populations in the energy eigenbasis $\{\ket{0}, \ket{\uparrow}, \ket{\downarrow}, \ket{2}\}$~\cite{Schulenborg2016Feb}. From now on, we will express operators in Liouville space and denote them as rounded kets, e.g., $\Ket{x}$ for operator $x$. The corresponding covectors will be denoted by rounded bras, $\Bra{x}\bullet = \Tr(x^\dagger \bullet)$.
In particular, the dot density operator is denoted $\Ket{\rho} = \sum_{j}P_j \Ket{j}$, with $\Ket{j} = \proj{j}$ for $j=0,\uparrow, \downarrow, 2$.
The evolution of the quantum dot state is then given by a standard rate equation 
\begin{equation}\label{master_eq}
    \partial_t \Ket{\rho} = W \Ket{\rho},
\end{equation}
where the evolution kernel $W$ is represented by a matrix whose off-diagonal elements are given by Fermi's golden rule,
\begin{equation}\label{W}
    W = \Gamma\begin{pmatrix}
        -2 \,  f_{\epsilon}^+ &  f_{\epsilon}^- &  f_{\epsilon}^- & 0 \\
         f_{\epsilon}^+ & - f_{\epsilon}^- - f_{U}^+ & 0 &  f_{U}^- \\
         f_{\epsilon}^+ & 0 & -f_{\epsilon}^- - f_{U}^+ &  f_{U}^- \\
        0 &  f_{U}^+ &  f_{U}^+ & -2 \,  f_{U}^-
    \end{pmatrix}.
\end{equation}
We have denoted $f_\epsilon^\pm = f(\pm\epsilon)$ and $f_U^\pm = f(\pm\epsilon \pm U)$, where $f$ is the Fermi-Dirac distribution for the reservoir, $f(E) = [1 + \exp(\beta E)]^{-1}$ since we take the electrochemical potential of the reservoir as energy reference, namely $\mu = 0$. The parameter $\beta = 1/\kB T$ represents the inverse temperature.

In such a simple single-reservoir case, the unique steady state  of the master equation \eqref{master_eq} corresponds to the equilibrium state of the dot with the reservoir and is given by the Gibbs state
\begin{equation}\label{rho_eq}
    \rhoeq = \frac{1}{Z}\e^{-\beta H},
\end{equation}
with the partition function $Z = \Tr[\exp(-\beta H)]$.

\subsection{Relaxation dynamics}\label{sec:relax}

For our study of the relaxation dynamics of the system, we start by an eigenmode decomposition of the kernel $W$ and identify the decay modes relevant to the Mpemba effect. By diagonalizing Eq.~\eqref{W}, we obtain
\begin{equation}\label{W diag}
    W = -\gamc \Ketbra{c}{c'} - \gamp \Ketbra{p}{p'} -\gamma_s \Ketbra{s}{s'},
\end{equation}
where we have identified four eigenmodes with eigenvalues $0,-\gamc,-\gams$, and $-\gamp$ (the equilibrium state contributing to $W$ with zero weight).
Since we are dealing with a dissipative evolution, $W$ is not Hermitian and the relation between the left and right eigenvectors is not straightforward. However, using a so-called fermionic duality \cite{Schulenborg2016Feb}, a dissipative symmetry fulfilled by $W$, compact analytical expressions can be found for all modes and rates and we can associate a physical meaning to each of these modes\footnote{The following analysis can be straightforwardly generalized to the two-reservoir case~\cite{Vanherck2017Mar,Monsel2022Jul} studied in Ref.~\cite{Chatterjee2023Aug}, in order to provide an understanding of the modes contributing to the decay dynamics.}.
The first eigenmode, with eigenvalue $0$, corresponds to the equilibrium state. The associated right eigenvector is $\Ket{\rhoeq} = \sum_j \Peq_j\Ket{j}$, where $\Peq_j$ is the probability of having the dot in state $j=0,\uparrow,\downarrow,2$ at equilibrium. Using Eq.~\eqref{rho_eq}, we get $\Peq_0 = 1/Z$, $\Peq_\uparrow=\Peq_\downarrow = \e^{-\beta\epsilon}/Z$ and $\Peq_2 = \e^{-\beta(2\epsilon + U)}/Z$. The associated left eigenvector is $\Bra{\one}$, with $\one$ representing the identity operator. This covector corresponds to the trace operation, $\Bra{\one}\bullet = \Tr(\bullet)$.

The second eigenmode is the charge mode, corresponding to the charge response of the system, with the decay rate
\begin{equation}\label{gamc}
    \gamc = \Gamma(f_\epsilon^+ + f_U^-).
\end{equation}
The associated right and left eigenmodes are
\begin{align}\label{charge}
    \Ket{c} &= \frac{1}{2}(-1)^N\left[\Ket{N} - \bNeq\Ket{\one}\right], \\\nonumber
    \Bra{c'} &= \Bra{N} - \Neq \Bra{\one},
\end{align}
where $N = N_\uparrow + N_\downarrow$ is the particle number operator and we have been able to write the modes in a compact way by introducing the average numbers of electrons in the dot for the equilibrium state $\rhoeq$, $\Neq = \Peq_\uparrow + \Peq_\downarrow + 2\Peq_2$, $\bNeq = \bPeq_\uparrow + \bPeq_\downarrow + 2\bPeq_2$, as well as the average numbers of electrons in the dot for the equilibrium state $\brhoeq$ of a fictitious system with inverted energies $\epsilon\rightarrow-\epsilon,U\rightarrow-U$. The occurrence of the latter so-called \textit{dual state} emerges from our use of the fermionic duality and is not only valuable for a compact writing of the results but also for a straightforward parameter analysis. In the following, we will refer to all quantities with inverted energies $\epsilon\rightarrow-\epsilon,U\rightarrow-U$ by an overbar.

The third eigenmode is the spin mode, corresponding to the spin response of the system, with the decay rate
\begin{equation}\label{gams}
    \gams = \Gamma(f_\epsilon^- + f_U^+).
\end{equation}
The associated right and left eigenmodes are
\begin{align}\label{spin}
    \Ket{s} &= \frac{1}{2}\left[\Ket{\uparrow} - \Ket{\downarrow}\right], \\\nonumber
    \Bra{s'} &= \Bra{\uparrow} - \Bra{\downarrow}.
\end{align}
Unsurprisingly, the spin mode will turn out not to contribute to the dynamics of the system as long as the spin symmetry is not broken by the preparation of the initial state.

Finally, the fourth mode is the parity mode, related to the odd/even parity of the number of electrons in the dot, with the decay rate
\begin{equation}\label{gamp}
    \gamp = 2\Gamma.
\end{equation}
The associated right and left eigenmodes are
\begin{align}\label{parity}
    \Ket{p} &= \Ket{(-\one)^N}, \\\nonumber
    \Bra{p'} &= \Bra{(-\one)^N\brhoeq},
\end{align}
with the parity operator $(-\one)^N$ and the equilibrium state of the fictitious dual system with inverted energies $\brhoeq$.
We now use the decay-mode decomposition of $W$ to, first, understand the relevant timescales and decay modes in the relaxation toward equilibrium of a dot prepared in an arbitrary initial state and, second, apply those results to the study of the Mpemba effect.

We consider the quantum dot prepared in some arbitrary initial state $\Ket{\rho(0)}$ at time $t = 0$ and then let it thermalize with the reservoir. Thanks to the eigenmode decomposition of the kernel, Eq.~\eqref{W diag}, we can easily express the dot state at time $t \ge 0$ as~\cite{Splettstoesser2010Apr,Schulenborg2016Feb}
\begin{equation}\label{rho(t)}
    \Ket{\rho(t)} = \Ket{\rhoeq} + c(t) \Ket{c} + p(t) \Ket{p} + s(t) \Ket{s},
\end{equation}
with the functions $c(t) = \Braket{c'}{\rho(0)}\e^{-\gamc t}$, $p(t) = \Braket{p'}{\rho(0)}\e^{-\gamp t}$, $s(t) = \Braket{s'}{\rho(0)}\e^{-\gams t}$ containing the overlaps between the initial state and the modes together with exponential factors for their decay. Since we consider the spin-degenerate case, namely $P_\uparrow = P_\downarrow$, we will always have $s = P_\uparrow - P_\downarrow = 0$ in the following [Eq.~\eqref{spin}]. Therefore, the spin decay mode never contributes to the relaxation here.
As there is no need to distinguish between the spin orientations, we define $P_1 =  P_\uparrow + P_\downarrow$ and introduce the occupation basis (describing all spin-degenerate states) $\{\Ket{0}, \Ket{1}, \Ket{2}\}$, with $\Ket{1} = \frac{1}{2}[\Ket{\uparrow} + \Ket{\downarrow}]$. The dot state can then be expressed as $\Ket{\rho} = \sum_{n=0}^2 P_n \Ket{n}$, where $P_n$ is the probability that the dot contains $n$ electrons.
The occupation basis is orthogonal, but not orthonormal since $\Braket{1}{1} = 1/2$. For convenience, we thus define the basis of covectors $\{\Bra{n'}\}$ with $\Bra{0'}=\Bra{0}$,  $\Bra{1'}=2\Bra{1}$ and $\Bra{2'}=\Bra{2}$ such that $P_n = \Braket{n'}{\rho}$ and $W = \sum_{l,n} \Mel{l'}{W}{n} \Ketbra{l}{n'}$ (see the Supplemental Material of Ref.~\cite{Schulenborg2016Feb} for details).

Note that by construction, any $\Ket{\rho}$ of the form given by Eq.~\eqref{rho(t)} has unit trace, $\Tr\rho = \Braket{\one}{\rho} = 1$, since $\Braket{\one}{\rhoeq} = \sum_n \Peq_n = 1$ and $\Braket{\one}{c} = \Braket{\one}{p} = 0$. However, not all values of $(c, p)$ give physical states, that is with positive populations $P_n = \Braket{n'}{\rho}$, $n= 0, 1, 2$. In the $(c, p)$-plane, all physical states $\Ket{\rho} = \Ket{\rhoeq} + c \Ket{c} + p \Ket{p}$ are contained in a triangle with vertices corresponding to the states $\Ket{0}, \Ket{1}, \Ket{2}$ (see proof in Appendix \ref{app:c,p}). This can be seen in Fig.~\ref{fig:Mpemba}, which will be analyzed in more detail in Section \ref{sec:Mpemba}.

In order to understand the Mpemba effect, it is crucial to know the relative magnitudes of the decay rates as well as which of the decay modes contribute to the initial state and which of them are visible in the detected observable.
For the decay rates, we find that $\gamc \leq \gamp$ [recall that $\gams$ does not contribute here and see Eqs.~\eqref{gamc} and \eqref{gamp}], see Figs.~\subfigref{fig:rates}{b-d}. This means that the charge mode is the slowest to decay, while the parity mode has an exponential decay at the larger rate $\gamp$.
Accordingly, we can possibly find a Mpemba effect depending on how much this faster mode is excited in the initial state. We even have an exponential speedup when the charge component, corresponding to the slowest decay mode, is fully suppressed in a specific initial state, namely $\Braket{c'}{\rho(0)} = 0$. How strong this exponential speedup is depends also on how different the decay rates $\gamc$ and $\gamp$ are.

Importantly, the relative magnitudes of the rates are strongly influenced by the strength and the sign of the interaction $U$, see Figs.~\subfigref{fig:rates}{b-d}. In the case of vanishing interactions, $\beta\abs{U} \ll 1$, we always have $\gamc \simeq \gamp/2$, and, therefore one can expect a significant speedup. This remains true also for the case of strong repulsive interactions, $\beta U \gg 1$, as long as $\epsilon$ is significantly smaller than $-U$ or significantly larger than 0 (with respect to the temperature energy scale $\kB T$), $\gamc \simeq \gamp/2$. In contrast, for $-U \lesssim \epsilon \lesssim 0$ in the Coulomb-blockaded region, one finds $\gamc\simeq\gamp$ such that both a possible Mpemba effect and exponential speedup become much less pronounced. This is decisively different for the case of attractive onsite interaction~\cite{Cheng2015May,Hamo2016Jul,Cheng2016Dec,Prawiroatmodjo2017Aug}, $-\beta U \gg 1$. For  $0 \lesssim \epsilon \lesssim \abs{U}$, $\gamc\ll \gamp$ becomes almost zero such that a possible exponential speedup would have a major effect. The underlying slowing down of the charge relaxation was identified as a limiting factor in the efficiency of driven thermal machines \cite{Monsel2022Jul, Monsel2023Dec}.

We furthermore need to consider whether we can find initial states showing the Mpemba effect and/or decaying with exponential speedup and the interaction also plays an important role in the state preparation. In the limits $\beta\epsilon, \beta(\epsilon + U) \gg 1$ and $\beta\epsilon, \beta(\epsilon + U)  \ll - 1$, the equilibrium state (respectively $\Ket{0}$ and $\Ket{2}$) becomes the only physical state with zero charge amplitude [the $c = 0$ line in the $(c,p)$-plane in Fig.~\ref{fig:Mpemba}] such that it is impossible to obtain an exponential speedup in the relaxation (see Appendix \ref{app:c,p}).
Whether the Mpemba effect is then indeed visible in a given observable does furthermore not only depend on the initially prepared state $\rho(0)$,  but also on the overlap between the observable and the decay modes of the quantum-dot state, as discussed in the following Secs.~\ref{sec:Mpemba_relax} and \ref{sec:Mpemba}.

\subsection{Decay of observables relevant to the Mpemba effect}\label{sec:Mpemba_relax}

Now with the quantum dot model and the decay dynamics of its state in place, we are in the position to set up the analysis of the time-dependent behavior of the observables of our interest.

\begin{figure*}[t]
    \includegraphics[width=\linewidth]{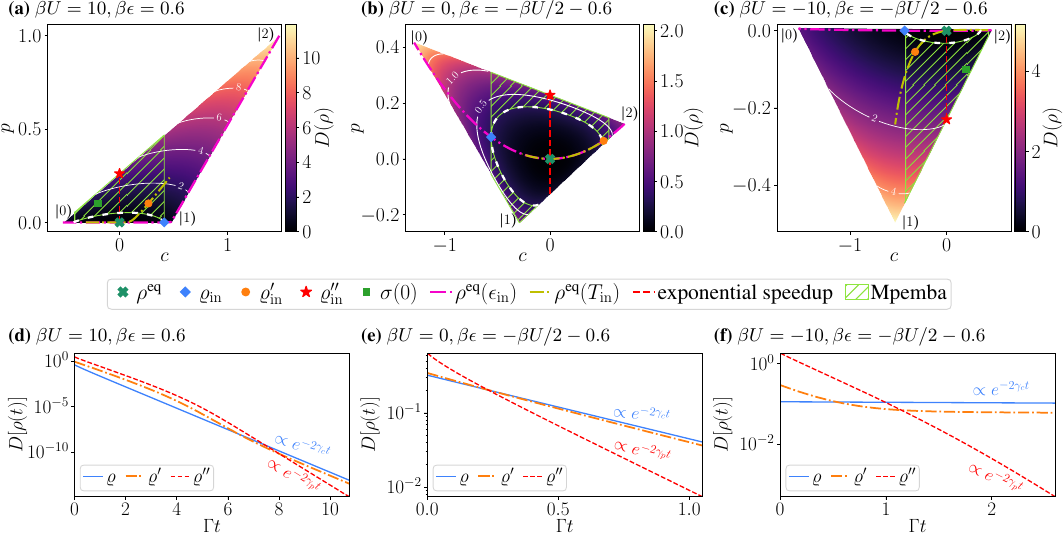}
    \caption{\label{fig:Mpemba}
        \textbf{(a)}, \textbf{(b)}, \textbf{(c)} Relative entropy $\Deq$ as a measure of the distance to equilibrium of all possible physical states, defined by their charge amplitude $c$ and parity amplitude $p$, i.e., $\Ket{\rho} = \Ket{\rhoeq} + c\Ket{c} + p\Ket{p}$,  for a given value of $\epsilon$ and $U$; in particular for {(a)} repulsive interaction, {(b)} no interaction and {(c)} attractive interaction. The area hashed in green corresponds to initial states that show a Mpemba effect compared to the initial state $\varrho_\ii$ (blue diamond). The dashed red line shows the initial states giving an exponential speed-up. The red star indicates an initial state $\varrho_\ii''$ exhibiting a strong Mpemba effect and the orange circle an initial state $\varrho_\ii'$ giving the Mpemba effect without exponential speedup, in both cases with respect to the initial state $\varrho_\ii$. The green cross corresponds to the equilibrium state at given $\epsilon$, $U$, and $T$. The green square marks the initial state of the fourth energy decay curve in Fig.~\ref{fig:Energy}. The magenta and yellow dash-dotted lines indicate all initial states that can be prepared by a gate-voltage switch and a temperature quench respectively, see Sec.~\ref{sec:init state}. \textbf{(d)}, \textbf{(e)}, \textbf{(f)} Time evolution of the relative entropy $\Deq$ during the relaxation from the initial states $\varrho_\ii$, $\varrho_\ii'$, and $\varrho_\ii''$ indicated in panels (a) to (c).
    }
\end{figure*}

We first consider the relative entropy, see Eq.~\eqref{def_Deq}. The relative entropy also determines the nonequilibrium free energy difference, which is defined in Eq.~\eqref{def DF} and which was used for the analysis of the Mpemba effect in previous works~\cite{Moroder2024Apr}. In Liouville space, the relative entropy $\Deq$ can be expressed as
\begin{align}\label{Drho}
    \Deq[\rho(t)] =\,& \Braket{\log\rho(t)}{\rho(t)} -  \Braket{\log\rhoeq}{\rho(t)}
\end{align}
with $\Bra{\log\rho} = \log (P_0) \Bra{0'} + \log(P_1/2) \Bra{1'}+ \log (P_2)\Bra{2'}$ (remember that the single-electron state is a mixed state).
Expanding each $\log[P_n(t)]$ into a Taylor series with respect to $\Braket{n'}{\rhoeq} =\Peq_n$, as shown in Appendix \ref{app:Deq}, and using the expression for the decay dynamics of the quantum state~\eqref{rho(t)}, we find
\begin{equation}\label{Deq(t)}
    \Deq[\rho(t)] = \sum_{k> 1} A_k  \sum_{j=0}^{k}B_{k,j}\,p(0)^j c(0)^{k-j}\e^{-\gamma_{k,j}t}.
\end{equation}
Here, we have defined
\begin{subequations}\label{ABg}
    \begin{align}
        A_k &= \frac{(-1)^{k}}{(k-1)k}, \label{A}\\
        B_{k,j} &=\frac{k!}{j!(k-j)!}\sum_{n=0,1,2}\frac{\Braket{n'}{p}^j\Braket{n'}{c}^{k-j}}{(\Peq_n)^{k-1}},\label{B}\\
        \gamma_{k,j} &=(k-j)\gamc +j\gamp,\label{g}
    \end{align}
\end{subequations}
where the $\gamma_{k,j}$ are combinations of multiples of the two decay rates $\gamc$ and $\gamp$. In general [$c(0)\neq0$], the relevant decay time-scale for the long-time-behavior of $\Deq$ is set by the smallest rate $\gamma_{2,0} = 2\gamc$ (note that $k$ only takes values $k>1$). Indeed, in the long-time limit, we have
\begin{equation}\label{Deq}
    \Deq[\rho(t)] \underset{\gamc t\to\infty}{\sim} \frac{Z\bar{Z}}{8}\bdNeq  c^2\e^{-2\gamc t},
\end{equation}
where $\bdNeq = \Braket{(N - \bNeq)^2}{\brhoeq}$ the variance of the particle number in the fictitious dual system with inverted energies and $\bar{Z}$ its partition function.

As a second observable, we study the evolution of the internal dot energy, $E[\rho(t)] = \Braket{H}{\rho(t)}$, which is shown in Fig.~\ref{fig:Energy}.
Using the expression for $\Ket{\rho(t)}$ from Eq.~\eqref{rho(t)}, we get
\begin{align}\label{E(t)}
    \Delta E[\rho(t)] &=  E[\rho(t)] - E_\text{eq}\\\nonumber
    & = E_c(0) \e^{-\gamc t} + E_p(0) \e^{-\gamp t},
\end{align}
where $E_c(0) = E_\text{Seebeck} c(0)$ and $E_p(0) = U p(0)$ are the energy contributions from the charge and parity components of the initial state $\rho(0)$. We have denoted $E_\text{Seebeck} = \epsilon + \frac{U}{2}(2 - \bNeq)$  the Seebeck energy of the dot~\cite{Monsel2022Jul}.
These identities clearly show that in order to observe the Mpemba effect in the dot-energy dynamics, electron-electron interaction is required. Otherwise, the decay of the internal energy of any state is completely governed by the rate $\gamma_c$, since the parity rate is only visible in observables containing correlation between all single-particle states of the system~\cite{Schulenborg2018}. In contrast, this occurrence of correlations is guaranteed in the relative entropy---involving a logarithm of the density operator---independently of the electron-electron interaction. Equivalently, this means that in order to see the Mpemba effect, we generally need to analyze observables that are not solely governed by the (slow) charge mode.

In Section \ref{sec:Mpemba}, we discuss the Mpemba effect based on the decay dynamics of these two quantities, Eqs.~\eqref{Drho} and \eqref{E(t)}.

\section{Mpemba effect in a quantum dot}\label{sec:Mpemba}

\subsection{Mpemba effect based on the nonequilibrium free energy}

Using the insights we gain from the analysis of the decay dynamics in Sec.~\ref{sec:Mpemba_relax}, we now study the Mpemba effect in the quantum dot as defined in Sec.~\ref{sec:def Mpemba} and \ref{sec:Mpemba_relax}.
To fulfill the first condition \eqref{Mpemba_1} for the Mpemba effect to occur, we choose two initial states, $\varrho_\ii$ and $\varrho_\ii'$, such that $D(\varrho_\ii') > D(\varrho_\ii)$, where the relative entropy is defined in Eq.~\eqref{def_Deq}. We denote respectively by $\varrho(t)$ and $\varrho'(t)$ the state of the dot during the relaxation from each initial state, as given by Eq.~\eqref{rho(t)} for $\varrho(0) = \varrho_\ii$ and $\varrho'(0) = \varrho_\ii'$.
To check the second condition \eqref{Mpemba_2} for the Mpemba effect, we use Eq.~\eqref{Deq(t)} to express the relative entropy difference $\Deq[\varrho'(t)] - \Deq[\varrho(t)]$, denoting the initial charge and parity amplitudes as $ c_\ii= \Braket{c'}{\varrho_\ii}$, $p_\ii = \Braket{p'}{\varrho_\ii}$ and $c_\ii' = \Braket{c'}{\varrho_\ii'}$, $p_\ii' = \Braket{p'}{\varrho_\ii'}$. We then see that in the long time limit, in general, namely for $c_{\ii}'^{2}\neq c_{\ii}^2$,
\begin{equation}\label{DF}
   \Deq[\varrho'(t)] - \Deq[\varrho(t)] \underset{{\gamc t\to\infty}}{\sim}\frac{Z\bar{Z}}{8}\bdNeq  (c_{\ii}'^2 - c_{\ii}^2)\e^{-2\gamc t}.
\end{equation}
Therefore, if $\abs{c_{\ii}'} < \abs{c_{\ii}}$, $\Deq[\varrho'(t)] - \Deq[\varrho(t)]$ will eventually become negative.
All in all, sufficient conditions to obtain the Mpemba effect are therefore
\begin{equation}\label{Mpemba condition}
   \Deq(\varrho_\ii') > \Deq(\varrho_\ii) \text{ and } \abs{c_{\ii}'} < \abs{c_{\ii}}.
\end{equation}
If, in addition, $c_{\ii}' = 0$, we get a strong Mpemba effect since the dynamics of $\varrho'(t)$ is governed by the parity decay rate $\gamp$ alone while the one of $\varrho(t)$ is dominated by the slower charge decay rate $\gamc$.

Note that even when both initial states have finite amplitudes for only one mode, namely $c_\ii = c_\ii' = 0$ or $p_\ii = p_\ii' = 0$, several rates $\gamma_{k,j}$ [defined in Eq.~\eqref{g}], namely multiples of the decay rate of the only relevant mode, contribute to the evolution of the relative entropy [Eq.~\eqref{Deq(t)}], due to the logarithm in the definition of $\Deq$  [Eq.~\eqref{def_Deq}]. This means that we can still get a Mpemba effect for such a pair of states that both relax with the same unique mode (charge or parity), though this is not the physically most interesting case since the crossover is uniquely due to the choice of the observable and not due to competing relaxation dynamics in the dot state. This means that for getting a physically meaningful Mpemba effect, we need in addition to Eq.~\eqref{Mpemba condition} a nonvanishing $p_\ii'$.

We have graphically represented the condition \eqref{Mpemba condition} in Fig.~\ref{fig:Mpemba} for different scenarios concerning the electron-electron interactions: strong repulsive interaction in panel (a), no interaction in panel (b) and strong attractive interaction in panel (c).
These plots show triangles containing all possible states $\rho$ for given values of $\epsilon, U$, and $T$ in the $(c, p)$-plane.
The shape of the triangle formed by all physical states depends strongly on the values of both $U$ and $\epsilon$. For instance, in Fig.~\subfigref{fig:Mpemba}{a}, the parity amplitudes are almost all positive while in Fig.~\subfigref{fig:Mpemba}{c} they are almost all negative. By a green cross, we mark the equilibrium state $\rhoeq$ toward which all other initial states relax for the indicated values of $\epsilon$ and $U$. For a strong interaction, $\beta\abs{U} \gg 1$, $\rhoeq$ is always on an edge of the triangle, like in Figs.~\subfigref{fig:Mpemba}{a} and \subfigref{fig:Mpemba}{c}, while for weaker interaction, it can be inside the triangle, like in Fig.~\subfigref{fig:Mpemba}{b} (see Appendix~\ref{app:c,p}). Also, the energy $\epsilon$ plays a role in the position of $\rhoeq$. In particular, when both transition energies, $\epsilon$, and $\epsilon + U$, are way above the reservoir electrochemical potential or way below it, $\Ket{\rhoeq}$ goes to $\Ket{0}$ or $\Ket{2}$ respectively, namely to one of the vertices of the triangle.

The plots in the upper row of Fig.~\ref{fig:Mpemba} show for each state $\rho$ the relative entropy with respect to the equilibrium state, $\Deq(\rho)$. The position of the equilibrium state, discussed above, influences the shape of the isolines of $\Deq$ (white lines in Fig.~\ref{fig:Mpemba}).
Now, to discuss the Mpemba effect, we choose a first ``colder'' initial state, $\varrho_\ii$, indicated by the blue diamond, with a relatively low relative entropy and a relatively large charge amplitude $\abs{c_\ii}$ in order to easily find ``hotter'' states $\varrho_\ii'$ fulfilling the Mpemba condition \eqref{Mpemba condition}.
Then, we show the isoline of the states with the same relative entropy with a white dashed line and draw vertical lines at $\pm c_\ii$ to visualize which states $\varrho_\ii'$ fulfill condition \eqref{Mpemba condition} (green hashed area). This means that all states in the green hashed area show the Mpemba effect compared to the reference initial state $\varrho_\ii$, when analyzing the relative entropy.

If in addition, the charge amplitude of an initial state is suppressed, there is an exponential speedup with respect to the initial state $\varrho_\ii$. All states fulfilling this condition are positioned on the vertical dashed red line. This also shows that, if the equilibrium is a pure state, $\Ket{0}$ or $\Ket{2}$, then it is the only physical state on the $c = 0$ line and, therefore, it is never possible to get an exponential speedup of the relaxation or a strong Mpemba effect.
By contrast, states that are both in the green hashed area and on the red dashed line result in the strong Mpemba effect with respect to $\varrho_\ii$.

To confirm these findings, we choose three different initial states, $\varrho_\ii$ (blue diamond and solid blue line), a state $\varrho_\ii'$ (orange dot and orange dash-dotted line) chosen to give the Mpemba effect without exponential speedup and a state $\varrho_\ii''$ (red star and dashed red line) giving the strong Mpemba effect, and show the time-evolution of their relative entropies $D(t)$ in Figs.~\subfigref{fig:Mpemba}{d-f}.
As expected from Eq.~\eqref{Deq}, $\Deq$ exhibits an exponential decay with rate $2\gamc$ in the long-time limit for $\varrho(t)$ and $\varrho'(t)$, but since $c_\ii'' = 0$, $\Deq[\varrho''(t)]$ decays with rate $2\gamp$.
Both orange and red curves eventually cross the blue curve, confirming the occurrence of the Mpemba effect. However, the time $t_\M$ at which the crossing occurs is strongly dependent on the type of electron-electron interaction. For strong repulsive interaction, the crossing happens much later than for vanishing or attractive interaction.
This can be understood from our discussion of the values of the charge decay rate in Sec.~\ref{sec:relax}. For the parameters chosen here, we have $\gamc/\gamp \simeq 0.68$ in Fig.~\subfigref{fig:Mpemba}{d}, $\gamc/\gamp = 0.5$ in Fig.~\subfigref{fig:Mpemba}{e} and $\gamc/\gamp \simeq 0.008$ in Fig.~\subfigref{fig:Mpemba}{f}. The larger the difference between the rates $\gamc$ and $\gamp$, the earlier the crossing between the relative entropies occurs. The exact time of the crossing also depends on how far apart the relative entropies of the initial states $\varrho_\ii$ and $\varrho_\ii'$ (or $\varrho_\ii''$) are. They are much closer together for $\beta U = 0$, see Figs.~\subfigref{fig:Mpemba}{b} and \subfigref{fig:Mpemba}{e}, than for $\beta U = -10$, see Figs.~\subfigref{fig:Mpemba}{c} and \subfigref{fig:Mpemba}{f}, which explains why the crossing happens earlier in the former case despite the smaller difference between rates.
Note that in Figs.~\subfigref{fig:Mpemba}{a} and \subfigref{fig:Mpemba}{d}, we even chose a value of $\epsilon$ that is not too close to the electron-hole symmetric point (unlike in the other panels), otherwise the difference between rates $\gamc$ and $\gamp$ would have been too small to see the Mpemba effect within a reasonable time range.

\subsection{Mpemba effect based on the energy}\label{sec:Mpemba energy}

\begin{figure}[!]
    \includegraphics[width=\linewidth]{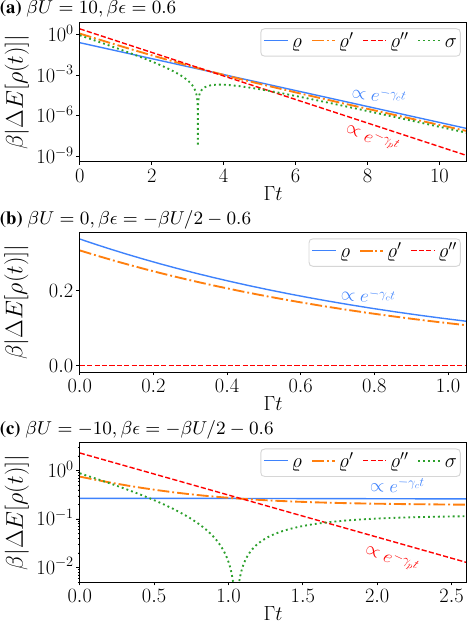}
    \caption{\label{fig:Energy}
        Time evolution of the energy distance to equilibrium, $\beta\,\abs{\Delta E[\rho(t)]}$, in the same cases as in Fig.~\ref{fig:Mpemba}. In addition, in panel (a), the green dotted line corresponds to the relaxation of the quantum-dot energy from an initial state with charge amplitude $\Braket{c'}{\sigma(0)} = -0.2$ and parity amplitude $\Braket{p'}{\sigma(0)} = 0.1$, indicated by a green square in Fig.~\subfigref{fig:Mpemba}{a}. In panel (c), the green dotted line corresponds to the relaxation of the quantum-dot energy from an initial state with charge amplitude $\Braket{c'}{\sigma(0)} = 0.2$ and parity amplitude $\Braket{p'}{\sigma(0)} = -0.1$, see the green square in Fig.~\subfigref{fig:Mpemba}{c}.
    }
\end{figure}

We now furthermore analyze the decay of the internal energy of the quantum dot, see Eq.~\eqref{E(t)}, for the three initial states $\varrho_\ii,\varrho_\ii',\varrho_\ii''$ indicated in Fig.~\ref{fig:Mpemba}.
For strong electron-electron interactions, we observe in Figs.~\subfigref{fig:Energy}{a} and \subfigref{fig:Energy}{c} that also the decay of the internal dot energy is faster for states with an initially higher energy if those states show the Mpemba effect in the relative entropy. This is evidenced by the crossings of the dashed red and the dash-dotted orange lines with the solid blue one. For the internal-energy decay in the case of attractive electron-electron interactions, the Mpemba effect occurs earlier and is more pronounced thanks to the large difference between the two involved decay rates $\gamc$ and $\gamp$. Note that the decay rates here occur with a factor differing by 2 in the exponents compared to the relative-entropy decay, which is due to the logarithmic dependence on $\rho$ in the relative entropy.
However, the parity contribution to the energy vanishes in the absence of electron-electron interaction, see Eq.~\eqref{E(t)}. This is especially visible for state $\varrho''$ in Fig.~\subfigref{fig:Energy}{b} since this state has no charge amplitude ($c_\ii'' = 0$) and $\Delta E[\varrho''(t)] = 0$ at all times. Hence, as explained in Sec.~\ref{sec:Mpemba_relax}, in the absence of interaction, the internal dot energy does not allow probing the parity mode and hence it does not allow to reveal the Mpemba effect either.

Though for strong interactions the states we chose exhibited the Mpemba effect both for $D$ and $\Delta E$, this is not in general the case since those two distance measures are not equivalent. Indeed, from Eq.~\eqref{E(t)} in the long time limit, for $c_\ii \neq c_\ii'$ and $E_\text{Seebeck} \neq 0$, we get
\begin{equation}
 \Delta E[\varrho'(t)] - \Delta E[\varrho(t)] \underset{{\gamc t\to\infty}}{\sim} (c_\ii' - c_\ii) E_\text{Seebeck} \e^{-\gamc t}.
\end{equation}
Therefore, a sufficient condition to get the Mpemba effect based on the energy is
\begin{equation}\label{Mpemba condition E}
   \Delta E(\varrho_\ii') > \Delta E(\varrho_\ii) \text{ and } \left|\begin{array}{ll}
       {c_{\ii}'} < {c_{\ii}} & \text{if }E_\text{Seebeck}> 0 \\
       {c_{\ii}'} > {c_{\ii}} & \text{if }E_\text{Seebeck}< 0
   \end{array}\right. ,
\end{equation}
which is clearly not equivalent to the condition \eqref{Mpemba condition} for the nonequilibrium free energy.

Another important difference in the property of the energy decay compared to the relative-entropy decay is that the energy decay is not necessarily monotonic. For strong interaction, the energy can cross the equilibrium energy during the relaxation, as shown for state $\sigma(t)$ (green dotted line) in Figs.~\subfigref{fig:Energy}{a} and \subfigref{fig:Energy}{c}.
From Eq.~\eqref{E(t)}, we see that this happens if $E_c(0)$ and $E_p(0)$ have different signs while $\abs{E_p(0)} > \abs{E_c(0)}$,
since the parity contribution decays faster. A detailed analysis of how the monotonicity of the energy current depends on the ratio of the decay mode amplitudes has been carried out in Ref.~\cite{Ortmanns2023Aug}.

In summary, we find that also the decay of the internal energy can be faster for quantum dots with an initially higher energy if the electron-electron interaction is finite.
For finite interaction strengths only, one can hence even define the Mpemba effect based on the internal dot energy, but compared to the relative entropy (or the nonequilibrium free energy) it has the disadvantage of not being monotonic. This drawback also occurs in the effective temperature $T(t) = \partial{E}/\partial{S_\text{vN}}$ used in Ref.~\cite{Chatterjee2023Aug}, where $S_\text{vN}$ denotes the von Neumann entropy of the dot.
For $U = 0$, instead, the energy decay reduces to the single-exponential $\Delta E[\rho(t)] = \epsilon c(0) \e^{-\gamc t}$, such that the Mpemba effect cannot occur.

\section{Preparation of the initial states}\label{sec:init state}

In this section, we propose two methods to prepare initial states $\varrho_\ii$ and $\varrho_\ii'$ suitable for observing the Mpemba effect, either by changing the single-level energy of the dot via a rapid gate switch or by changing the temperature of the bath at time $t = 0$. For completeness, we also consider changing the interaction energy $U$ and preparing a nonequilibrium steady state with two reservoirs in Appendix~\ref{app:prep}. In this section, we will solely discuss the case of the relative entropy $D$ as a measure of the distance to equilibrium since we saw in Sec.~\ref{sec:Mpemba} that it gives a Mpemba effect for all interaction strengths, unlike the energy.

\subsection{Gate-voltage switch}\label{sec:energy quench}

\begin{figure*}[t!]
    \includegraphics[width=\linewidth]{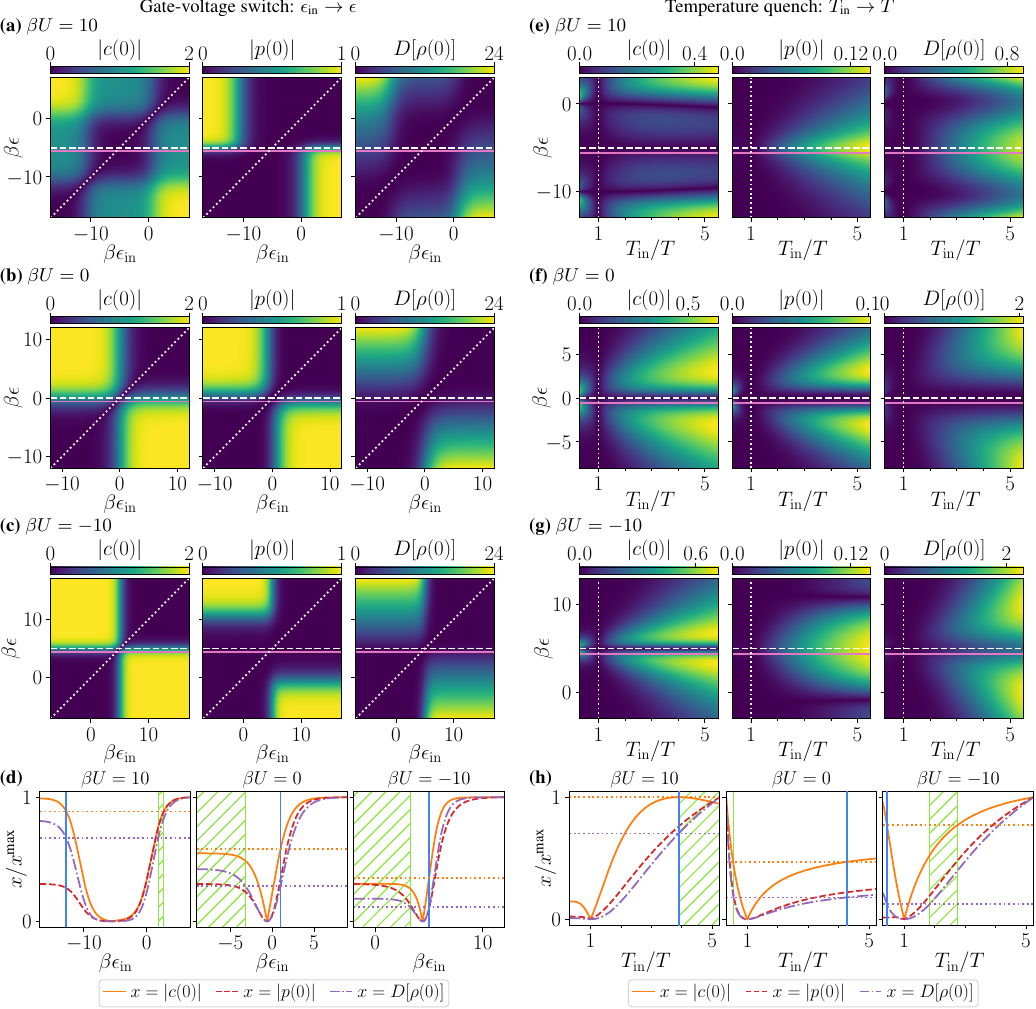}\vspace{-0.25cm}
    \caption{\label{fig:energy quench}\label{fig:T quench}
        \textit{Left column:} Gate-voltage switch as state preparation for the Mpemba effect.
        The absolute values of the charge amplitude and parity amplitude, as well as the relative entropy (with respect to the equilibrium state at energy $\epsilon$ and temperature $T$) of the prepared initial state $\rho(0) = \rhoeq(\epsilon_\ii)$ are plotted as functions of $\epsilon_\ii$ and $\epsilon$ for \textbf{(a)} repulsive interaction, \textbf{(b)} no interaction, and \textbf{(c)} attractive interaction. The dotted white line corresponds to the respective values for $\rho(0) = \rhoeq(\epsilon)$ and the dashed white line indicates the electron-hole symmetric point $\epsilon = -U/2$. \textbf{(d)} Slices taken from (a), (b), and (c) at the energy indicated by the solid pink lines, $\beta\epsilon = -\beta U/2  - 0.6$. All quantities are normalized by their maximum. The vertical blue lines indicate the energy values $\varepsilon_\ii$ used to prepare $\varrho_\ii$. The areas hashed in green correspond to the range of values of $\varepsilon_{\ii}'$ that yield an initial state $\varrho'_\ii$ showing a Mpemba effect compared to the initial state $\varrho_\ii$.
        \textit{Right column:} Temperature quench as state preparation for the Mpemba effect.
        The absolute values of the charge amplitude and parity amplitude, as well as the relative entropy (with respect to the equilibrium state at energy $\epsilon$ and temperature $T$) of the prepared initial state $\rho(0) = \rhoeq(T_\ii)$ are plotted as functions of $T_\ii$ and $\epsilon$ for \textbf{(e)} repulsive interaction, \textbf{(f)} no interaction and \textbf{(g)} attractive interaction. The dotted white line marks the values corresponding to $\rho(0) = \rhoeq(T)$ and the dashed white line indicates the electron-hole symmetric point $\epsilon = -U/2$. \textbf{(h)} Slices taken from (e), (f), and (g) at the energy highlighted by the solid pink lines, $\beta \epsilon = -\beta U/2  - 0.6$. The vertical blue line marks the temperature $\mathcal{T}_\ii$ used to prepare $\varrho_\ii$. The area hashed in green corresponds to the range of values of $\mathcal{T}_\ii'$ suitable to prepare an initial state $\varrho'_\ii$ showing a Mpemba effect compared to the initial state $\varrho_\ii$. \textit{Both columns:} In panels (d) and (h), the dotted orange and purple lines are guides to the eye indicating $\abs{c_{\ii}}$ and $\Deq(\varrho_\ii)$ respectively, that is, the values of $\abs{c(0)}$ and $D[\rho(0)]$ at the vertical blue line, to make apparent that the area in green fulfills the Mpemba conditions \eqref{Mpemba condition}.
        \vspace{-0.25cm}
    }
\end{figure*}

We start by considering a switch of the single-particle energy of the dot, which can be achieved experimentally by switching a gate voltage. The preparation adheres to the following procedure: First, the dot is prepared in its equilibrium state for energy $\epsilon_\ii$, then, at time $t=0$, the dot energy is instantaneously set to a new value $\epsilon$.
Thereby, the prepared initial state is $\rho(0) = \rhoeq(\epsilon_\ii) = \exp(-\beta H_{\epsilon_\ii})/Z_{\epsilon_\ii}$ [Eq.~\eqref{rho_eq}], where $H_{\epsilon_\ii}$ is the dot Hamiltonian as given in Eq.~\eqref{H_dot}, with the level energy $\epsilon_\ii$ and the partition function $Z_{\epsilon_\ii}$ corresponding to the values before the switch. After the switch, this state represents a non-equilibrium probability distribution of populations in the quantum dot with energy $\epsilon$.
In the left column of Fig.~\ref{fig:energy quench}, we show the charge amplitude, parity amplitude, and relative entropy of this initial state as functions of $\epsilon_\ii$ and $\epsilon$.

The plots in Figs.~\subfigref{fig:energy quench}{a-c} show features occurring at the electron-hole symmetric points and features of Coulomb blockade, which can be understood in detail by exploiting fermionic duality, see Refs.~\cite{Schulenborg2016Feb,Schulenborg2018,Monsel2022Jul} for further discussions. Here, we instead focus on identifying regions in which a switch provides the property $|c_\ii'|<|c_\ii|$ while $D(\varrho_\ii')>D(\varrho_\ii)$ (at finite parity amplitude). One needs to keep in mind that the final value of the energy $\epsilon$, and the reservoir temperature $T$ have to be the same for $\varrho_\ii$ and $\varrho_\ii'$. Otherwise, the two initial states would not relax towards the same equilibrium state. In Figs.~\subfigref{fig:energy quench}{a-c} [and also in Figs.~\subfigref{fig:energy quench}{e-g}], this condition means that the pair of states we want to test for the Mpemba effect has to lie on a horizontal line of constant $\epsilon$ in every plot.

For all interaction strengths, the final $\epsilon$ should be taken close to\footnote{By ``close to'' we always mean compared to temperature, namely $\beta\abs{-U/2 - \epsilon} \lesssim 1$.}  the particle-hole symmetry point, $-U/2$, to prepare two states $\varrho_\ii = \rhoeq(\varepsilon_\ii)$ and $\varrho_\ii'= \rhoeq(\varepsilon_\ii')$ giving a Mpemba effect\footnote{As previously, we use calligraphic symbols, here $\varrho$ and $\varepsilon$, to denote specific choices of states or values interesting for the Mpemba effect.}. The reason for this is that for a final energy value $\epsilon$ outside the Coulomb-blockaded region, $c$, $p$, and $D(\rho)$ are monotonic as a function of $\epsilon_\ii$ with the same trend (all either increasing or all decreasing), meaning that the conditions~\eqref{Mpemba condition} cannot be met. We hence choose $\beta\epsilon = -\beta U/2  - 0.6$ in Fig.~\subfigref{fig:energy quench}{d}, indicating our choice for $\epsilon$ by horizontal pink lines in Figs.~\subfigref{fig:energy quench}{a-c}. Only for strong repulsive interaction, $\epsilon$ could have been chosen in a wider range of values, from close to $-U$ to close to 0.

The vertical blue line in Fig.~\subfigref{fig:energy quench}{d} marks our choice of energy $\varepsilon_\ii$ to prepare $\varrho_\ii$, which is motivated by the relatively large $c_\ii$ and the relatively low $D(\varrho_\ii)$ for that state. As a next step, a state $\varrho_\ii'$ has to be prepared that shows the Mpemba effect compared to $\varrho_\ii$. The intersections of the blue line with $\abs{c(0)}$ and $\Deq[\rho(0)]$, that is $\abs{c_{\ii}}$ and $\Deq(\varrho_\ii)$, define the maximal charge amplitude and minimal relative entropy the second state $\varrho_\ii'$ may have by the Mpemba conditions \eqref{Mpemba condition} (dotted orange and purple lines).
For our specific choice of $\varrho_\ii$, the green hashed area highlights the range of values of $\varepsilon_\ii'$ allowing for that.
Note that the states $\varrho_\ii$ and $\varrho_\ii'$ discussed here are not the same as the ones represented in Fig.~\ref{fig:Mpemba}, which will instead be discussed in Sec.~\ref{sec:combined} below.

Which initial states indeed show the Mpemba effect strongly depends on the sign and strength of the onsite interaction.
No interaction and strong attractive interaction, visualized in the rightmost two plots in Fig.~\subfigref{fig:energy quench}{d}, give similar results: $\varepsilon_\ii'$ can take any value smaller than a threshold value that is determined by our choice of $\varepsilon_\ii$. This works for any $\varepsilon_\ii$ larger than $-U/2$ such that  $\Deq(\varrho_\ii) < \lim_{\varepsilon_\ii'\to-\infty}\Deq(\varrho_\ii')$ and $\abs{c_{\ii}} > \lim_{\varepsilon_\ii'\to-\infty}\abs{c'_{\ii}}$. The possible values of $\varepsilon_\ii'$ are more restricted for strong repulsive interaction, and also the rate ratio $\gamc/\gamp$ is relatively close to 1 in this case, making the Mpemba effect hard to observe, see the discussion of Fig.~\subfigref{fig:Mpemba}{d}. Note, however, that for strong repulsive interaction, as mentioned above, a different choice of $\epsilon$ would have been possible, improving the rate ratio.\clearpage

Finally, by comparing the plots for the charge and parity amplitude, we can see that this method of applying a gate switch does not allow for preparing an initial state giving an exponential speedup, namely with $c(0) = 0$ but $p(0) \neq 0$. Indeed, the only state allowing achieving  $c(0) = 0$ also has $p(0) = 0$ and is therefore the equilibrium state.

\subsection{Temperature quench}\label{sec:temperature quench}

We now keep the energy of the dot fixed and instead consider that the dot is initially at equilibrium with a reservoir at the temperature $T_\ii$.
Then, at time $t=0$, the dot is decoupled from this reservoir and coupled to another reservoir at a different temperature $T$. Thus, the prepared initial state is $\rho(0) = \rhoeq(T_\ii) = \exp(-\beta_\ii H)/Z_{T_\ii}$, where $Z_{T_\ii} = \Tr[\exp(-\beta_\ii H)]$ is the corresponding partition function. This method to prepare the initial state is particularly interesting since we can associate actual temperatures to the prepared initial states, like in the classical Mpemba effect~\cite{Lu2017May}.
In the right column of Fig.~\ref{fig:T quench}, we show the charge amplitude, parity amplitude, and relative entropy of such initial states as functions of $T_\ii$ and $\epsilon$.

First, by comparing the plots for the charge and parity amplitude in Figs.~\subfigref{fig:T quench}{e-g}, we see that this method allows for preparing an initial state giving an exponential speedup [by fulfilling $c(0) = 0$ but $p(0) \neq 0$] for strong interactions, $\beta \abs{U} \gtrsim 1$, when choosing $\epsilon = -U/2$. Indeed, at the particle-hole symmetry point, the Fermi functions $f^\pm_\epsilon = f^\mp_U$ have the same value irrespective of temperature, and therefore the particle number $N=\Neq$ is temperature-independent, ensuring that $c = 0$, see Eq.~\eqref{charge}.
However, for a weak interaction, $p$ is also almost zero at this point, therefore $\Ket{\rho(0)} \simeq \Ket{\rhoeq}$, which is not the case for stronger interactions.
For a strong repulsive interaction, being at the particle-hole symmetry point means that $\gamc \simeq \gamp$ [see Fig.~\subfigref{fig:rates}{b}], such that there is not really any speedup, but values of $T_\ii$ for which the charge amplitude vanishes can also be found for $\epsilon \simeq 0$ and $\epsilon \simeq -U$. These yield more favorable rate ratios $\gamc/\gamp$.

Second, as for the gate-voltage switch, taking $\epsilon$ close to $-U/2$ is a good choice to prepare two states $\varrho_\ii = \rhoeq(\mathcal{T}_\ii)$ and $\varrho_\ii'= \rhoeq(\mathcal{T}_\ii')$ giving a Mpemba effect, as shown in Fig.~\subfigref{fig:T quench}{h}. However, weak interaction allows for only a very restricted range of possible values of $\mathcal{T}_\ii'$, while strong attractive interaction allows for a larger range. For a strong repulsive interaction, $\abs{c_\ii}$ has a maximum, which makes it possible to choose any temperature of $\mathcal{T}_\ii'$ larger than $\mathcal{T}_\ii$ when taking $\mathcal{T}_\ii$ to be at this maximum point (blue line in the leftmost panel).
In this specific case, we indeed obtain the Mpemba effect in a way closest to the originally observed \cite{Mpemba1969May} effect since we can associate an actual temperature, larger than the final temperature, to both initial dot states, and the colder state relaxes slower than the hotter one. Note that this is not the case for a weak interaction or a strong attractive interaction, where we prepare one of the initial states with a temperature lower than the temperature of the final equilibrium state. As for the gate-voltage switch, the observation of the Mpemba effect in the strong repulsive interaction case is hindered by the small difference between rates $\gamc$ and $\gamp$. Similarly here, the rate ratio can be brought farther away from 1 by changing the value of $\epsilon$ since $\abs{c(0)}$ exhibits a local maximum,  allowing finding temperatures such that $\mathcal{T}_\ii' > \mathcal{T}_\ii > T$ to prepare the initial states $\varrho_\ii$ and $\varrho_\ii'$, for any $\epsilon$ within the Coulomb-blockaded region, $-U<\epsilon<0$.

\subsection{Preparing states by combining different quenches}\label{sec:combined}
In this section, we shed light on possibilities to experimentally prepare the pairs of states showing a Mpemba effect which are suggested in Fig.~\ref{fig:Mpemba}.
It is not always possible to prepare a pair of interesting states with a quench in a \textit{single} parameter, only. Assume, for example, a strong repulsive interaction. As we have seen in Secs.~\ref{sec:energy quench} and \ref{sec:temperature quench}, the preparation of two states showing the Mpemba effect via a gate-voltage or a temperature quench alone only works for $\epsilon$ in the Coulomb-blockaded region. However, close to the electron-hole symmetric point, that is, in the case that was studied in Sec.~\ref{sec:energy quench}, the two involved decay rates are very similar, $\gamc \simeq \gamp$ [see Fig.~\subfigref{fig:rates}{b}], such that $t_\M\Gamma \gg 1$. Hence, experimental observation would be unrealistic. In Fig.~\subfigref{fig:Mpemba}{a}, we therefore suggest states involving an $\epsilon$ that is sufficiently different from the electron-hole symmetric point.

The question arises of how to prepare the states in Fig.~\subfigref{fig:Mpemba}{a} with realistically tunable parameters. One straightforward solution based on the discussion in the previous two sections is to prepare the states by combinations of quenches of the single-level energy and temperature. Specifically, for the case shown in Fig.~\subfigref{fig:Mpemba}{a}, we can prepare the state $\varrho_\ii$ by a switch in the single-level energy, and the state $\varrho_\ii'$ by a quench in the temperature, for example with $\beta\varepsilon_\ii = -2$ and $\mathcal{T}_\ii'/T = 10$. To illustrate this, we indicate with dash-dotted lines in Figs.~\subfigref{fig:Mpemba}{a-c} all the states that can be prepared by a gate-voltage switch (in magenta) and by a temperature quench (in yellow). Note that simultaneous switches in gate voltage and temperature lead to an extended space of accessible initial states, as shown in Appendix~\ref{app:simultaneous}.

In the noninteracting case, in Fig.~\subfigref{fig:Mpemba}{b}, both states $\varrho_\ii$ and $\varrho_\ii'$ can be prepared with gate-voltage switches, using $\beta \varepsilon_{\ii} = 0.55$ and $\beta\varepsilon_{\ii}' = -2.2$, but $\varrho_\ii'$ could also be prepared with a temperature quench (the yellow and magenta lines in Fig.~\subfigref{fig:Mpemba}{b} partially lie on top of each other).
For a strong attractive interaction, in Fig.~\subfigref{fig:Mpemba}{c}, it is in principle possible to prepare a pair of states giving the Mpemba effect with gate-voltage switches, for example by choosing $\varrho_\ii'$ on the magenta dash-dotted line but in between the equilibrium state and the state $|2)$. However, for this choice, the state $\varrho_\ii'$ has a relative entropy close to $\Deq(\varrho_\ii)$. We therefore choose a state with a relative entropy $\Deq(\varrho_\ii')$ farther away from $\Deq(\varrho_\ii)$ in Fig.~\subfigref{fig:Mpemba}{c} instead which can be prepared using a temperature quench ($\beta\varepsilon_\ii =-\beta U/2 -0.1$ and $\mathcal{T}_\ii'/T = 3$).

Finally, we see in all panels of Fig.~\ref{fig:Mpemba} that it is not possible to prepare an initial state $\varrho_\ii''$ giving the strong Mpemba effect with any of the methods discussed until here. However, already a combination of simultaneous gate-voltage and temperature switches would make this possible, see Appendix~\ref{app:simultaneous}.
Also, as explained in Sec.~\ref{sec:temperature quench}, for strong interactions we can prepare a state $\varrho_\ii''$ with $c_\ii'' = 0$ using a temperature quench at $\epsilon = -U/2$, and a state $\varrho_\ii$ with $c_\ii \neq 0$ and $D(\varrho_\ii) < D(\varrho_\ii'')$ using a gate-voltage switch, resulting in the strong Mpemba effect. As explained above, this would be easier to observe for a strong attractive interaction rather than a strong repulsive interaction due to the rate ratio $\gamc/\gamp$.

Further methods how to prepare states by modulating $U$ or by connecting to contacts at different biases are explained in Appendix~\ref{app:prep}.

\section{Conclusion}

In summary, we have provided a detailed analysis of the Mpemba effect in an interacting quantum dot utilizing thermodynamic measures. Quantum dots are of special interest here because of their simplicity yet intriguing decay dynamics. Based on an analytic treatment of the master-equation dynamics, we have been able to identify the characteristics that a set of quantum-dot states requires in order to display the Mpemba effect. In particular, we have shown that the value of electron-electron interaction plays an important role in how pronounced the Mpemba effect is and at which time the initially ``hotter'' system becomes ``colder'' than the initially ``colder'' one. We have shown that attractive interaction exhibits a particularly pronounced Mpemba effect since the two relevant decay rates---the charge and the parity rate---are very different in the Coulomb-blockaded region. As the main quantity to reveal the Mpemba effect, we have chosen the relative entropy (or free-energy difference); however, in addition, we have also evidenced that the dynamics of the internal energy stored on the dot shows the Mpemba effect, as long as the electron-electron interaction on the quantum dot is finite.
We have furthermore provided experimentally relevant protocols to prepare states exhibiting the Mpemba effect, one of them being to create initial states by connecting them to thermal baths at different temperatures.

This paves the way for future experiments for detecting the Mpemba effect via the experimentally accessible energy decay following the preparation of states, e.g. by gate and temperature switches.

\acknowledgments
We thank John Goold for inspiring discussions and for pointing us to the Mpemba effect. We also thank Ludovico Tesser and Henning Kirchberg for useful comments on the manuscript.
We acknowledge financial support from the Knut and Alice Wallenberg foundation through the fellowship program and from the European Research Council (ERC) under the European Union’s Horizon Europe research and innovation program (101088169/NanoRecycle). J. G. acknowledges support from CRC 1277 (project B09) during her stay at Chalmers.

\appendix

\section{Charge and parity amplitudes}\label{app:c,p}

The vertices of the triangle of physical states in the $(c,p)$-plane in Fig.~\ref{fig:Mpemba} correspond to the states $\Ket{0}, \Ket{1}, \Ket{2}$. Indeed, denoting $c_n = \Braket{c'}{n}$ and $p_n = \Braket{p'}{n}$ the coordinates of $\Ket{n}$, $n = 0,1,2$, in the $(c,p)$-plane, the point with coordinates $(c, p)$ is within the triangle if and only if the three following inequalities hold \cite{Coxeter1995Sep}:
\begin{subequations}\label{triangle}
    \begin{align}
    (c_0 - c)(p_1 - p) - (c_1 - c)(p_0 - p) \ge 0,\label{T1}\\
    (c_1 - c)(p_2 - p) - (c_2 - c)(p_1 - p) \ge 0,\label{T2}\\
    (c_2 - c)(p_0 - p) - (c_0 - c)(p_2 - p) \ge 0\label{T3}.
\end{align}
\end{subequations}
We now construct $\Ket{\rho} = \Ket{\rhoeq} + c\Ket{c} + p\Ket{p} = \sum_{n = 0,1,2}P_n\Ket{n}$. Here, $\Ket{\rho}$ represents a physical state if and only if $\forall n\; P_n \ge 0$, since $\Braket{\one}{\rho}=1$ is already guaranteed by construction.
Using $c_n = n -\Neq$ and $p_0 = \bPeq_0$,  $p_1 = -\bPeq_1/2$,  $p_2 = \bPeq_2$, we can express $c$ and $p$ as
\begin{align}
    c
      &=  P_1 + 2 P_2-\Neq,\\\nonumber
    p
    &= \bPeq_0 P_0 -\frac{1}{2} \bPeq_1 P_1 +  \bPeq_2 P_2.
\end{align}
Using these expressions and $\sum_n P_n = \sum_n \Peq_n= \sum_n \bPeq_n = 1$, we find that the left hand sides of inequalities \eqref{T1}, \eqref{T2}, \eqref{T3} respectively become $P_2$, $P_0$, and $P_1$. This proves that the $(c,p)$ amplitudes defining physical states of the dot are the ones contained in the triangle of vertices $\Ket{0}, \Ket{1}, \Ket{2}$.

\subsection{Exponential speedup and pure equilibrium state}\label{app:pure eq state}

If the equilibrium state is a pure state, i.e. $\Ket{\rhoeq} = \Ket{0}$, which occurs in the limit $\beta\epsilon, \beta(\epsilon + U) \gg 1$, or $\Ket{\rhoeq} = \Ket{2}$, in the limit $\beta\epsilon, \beta(\epsilon + U) \ll -1$, then the only physical value of $p$ when $c = 0$ is $p=0$.
Indeed, putting $c_0 = p_0 = c = 0$ [$\Ket{\rhoeq} = \Ket{0}$] in inequalities \eqref{T1} and \eqref{T3} gives $c_1 p\ge 0$ and $-c_2 p \ge 0$, with $c_1 = 1$ and $c_2 = 2$ in this case. Similarly, $c_2 = p_2 = c = 0$ [$\Ket{\rhoeq} = \Ket{2}$] in inequalities \eqref{T2} and \eqref{T3} gives $-c_1 p \ge 0$ and $c_0 p \ge 0$, with $c_0 = -2$ and $c_1 = -1$.
Therefore, in these two cases, we always have $p=0$ and it is hence impossible to get an exponential speedup since the only physical state on the $c=0$ line is the equilibrium state. Note that there is no such restriction for $\Ket{\rhoeq} = \Ket{1}$, which occurs in the limit $\beta \epsilon \ll -1$, $\beta (\epsilon + U)\gg 1$ (only possible for strong repulsive interaction).

\subsection{Position of the equilibrium state and dual equilibrium state}\label{app:eq}
When one of the inequalities in \eqref{triangle} is an equality, it means that $\Ket{\rho}$ is on the corresponding edge of the triangle. Furthermore, we have seen above that the left-hand sides of inequalities \eqref{T1}, \eqref{T2}, and \eqref{T3} respectively correspond to $P_2$, $P_0$, and $P_1$.
Therefore, if $P_2 = 0$, then $\Ket{\rho}$ is on the edge linking $\Ket{0}$ and $\Ket{1}$. Similarly, $P_0 = 0$ means that $\Ket{\rho}$ is on the edge linking $\Ket{1}$ and $\Ket{2}$ and $P_1 = 0$ means that $\Ket{\rho}$ is on the edge linking $\Ket{0}$ and $\Ket{2}$. We now apply this insight in order to determine the positions of the equilibrium state and dual equilibrium state within the triangle.

In the limit of large repulsive interaction, $\beta U \gg 1$, $\bPeq_1 \simeq 0$ for any value of $\epsilon$, meaning that $\Ket{\brhoeq}$ is on the edge of the triangle linking $\Ket{0}$ and $\Ket{2}$. On the contrary, $\Ket{\rhoeq}$ is on the edge of the triangle linking $\Ket{0}$ and $\Ket{1}$ if $\epsilon > -U/2$ or on the edge linking $\Ket{1}$ and $\Ket{2}$ if $\epsilon < -U/2$, see pink dash-dotted line in Fig.~\subfigref{fig:Mpemba}{a}.

In the limit of large attractive interaction, $-\beta U \gg 1$, we have the exact opposite, with $\Ket{\rhoeq}$ on the edge linking $\Ket{0}$ and $\Ket{2}$, see pink dash-dotted line in Fig.~\subfigref{fig:Mpemba}{c}, and $\Ket{\brhoeq}$ on the edge of the triangle linking $\Ket{0}$ and $\Ket{1}$ if $\epsilon > -U/2$ or on the edge of the triangle linking $\Ket{1}$ and $\Ket{2}$ if $\epsilon < -U/2$.

When the interaction strength becomes of the order of the temperature or smaller, $\beta\abs{U} \lesssim 1$, the previous arguments no longer hold and both $\Ket{\rhoeq}$ and $\Ket{\brhoeq}$ are inside the triangle when the dot energy is close to the electron-hole symmetric point, $\epsilon \simeq -U/2$, see pink dash-dotted line in Fig.~\subfigref{fig:Mpemba}{b}.

\section{Relative entropy in the long time limit}\label{app:Deq}
In this Appendix, we analyze the behavior of the relative entropy with respect to the equilibrium state at time $t$, $\Deq[\rho(t)]$, in the long time limit, namely $t \gg 1/\gamc, 1/\gamp$.

\subsection{Time-evolution of the relative entropy}

First, we compute $\Deq[\rho(t)]$ for any time $t > 0$ for a relaxation from an arbitrary physical initial state $\rho(0)$.
From the master equation \eqref{master_eq} and the kernel eigenmode decomposition [Eq.~\eqref{W diag}], we can write the state of the dot at any time using Eq.~\eqref{rho(t)}. Then, the relative entropy can be expressed as in Eq.~\eqref{Drho}. We now write the Taylor series for each $\log[P_n(t)]$, which gives
\begin{widetext}
    \begin{equation}
        \Bra{\log\rho(t)} =\Bra{\log\rhoeq }  + \sum_{k> 0} \frac{(-1)^{k+1}}{k}\sum_{n=0}^2  \left[\frac{\Braket{n'}{c}c(0)\e^{-\gamc t}+\Braket{n'}{p}p(0)\e^{-\gamp t}}{\Braket{n'}{\rhoeq}}\right]^k \Bra{n'},
    \end{equation}
    with $\Braket{n'}{\rhoeq} =\Peq_n$. And then, from Eq.~\eqref{Drho}, we get
    \begin{align}
        \Deq[\rho(t)]
        &= \sum_{k> 0} \!\frac{(-1)^{k+1}}{k}\!\!\sum_{n}\! \left[\frac{\Braket{n'}{c}c(0)\e^{-\gamc t}+\Braket{n'}{p}p(0)\e^{-\gamp t}}{\Peq_n}\right]^k\!\!\!P_n(t)\\\nonumber
        &=\sum_{k> 0} \!\frac{(-1)^{k+1}}{k}\!\!\!\sum_{n}\! \frac{\left[\Braket{n'}{c}c(0)\e^{-\gamc t}+\Braket{n'}{p}p(0)\e^{-\gamp t}\right]^k}{(\Peq_n)^{k-1}} +\sum_{k> 0} \!\frac{(-1)^{k+1}}{k}\!\!\!\sum_{n}\! \frac{\left[\Braket{n'}{c}c(0)\e^{-\gamc t}+\Braket{n'}{p}p(0)\e^{-\gamp t}\right]^{k+1}}{(\Peq_n)^k}\\\nonumber
        &=\sum_{k> 1} A_k\sum_{n}\! \frac{\left[\Braket{n'}{c}c(0)\e^{-\gamc t}+\Braket{n'}{p}p(0)\e^{-\gamp t}\right]^{k}}{(\Peq_n)^{k-1}},
    \end{align}
\end{widetext}
where $A_k$ is defined in Eq.~\eqref{A}.
To go from the first to the second line, we used the expression of $P_n(t)$ obtained from Eq.~\eqref{rho(t)}. Then, on the second line, the term $k = 1$ in the first sum is zero because $\Braket{\one}{c}=\Braket{\one}{p} = 0$, so we can merge the two sums by re-indexing the second one. Finally, we obtain Eq.~\eqref{Deq(t)} by binomial expansion.

\subsection{Exponential speedup}

From Eq.~\eqref{Deq(t)}, we can now study the time scales of the relaxation toward equilibrium. We always have $\gamc < \gamp$, though $\gamc \to \gamp$ when $\epsilon \to -U/2$ and $U \to +\infty$. The slowest decay rate is then $\gamma_{2,0} = 2\gamc$, and in the long time limit $\gamma_{2,0} t\gg 1$,
\begin{equation}
    \Deq[\rho(t)] \underset{\gamc t\to\infty}{\sim} \frac{B_{2,0} }{2}  c(0)^2\e^{-2\gamc t},
\end{equation}
provided that the initial state has a finite overlap with the charge mode, $c(0) \neq 0$. Using the expression of $\Ket{c}$ from Eq.~\eqref{charge} and \cite{Schulenborg2018Dec}
\begin{equation}\label{bP_P}
    \bPeq_n = \frac{(\Peq_n)^{-1}}{\sum_{m}(\Peq_m)^{-1}}= \frac{1}{Z\bar{Z}\Peq_n},
\end{equation}
the coefficient $B_{2,0}$ can be expressed as
\begin{equation}\label{B20}
    B_{2,0} = \frac{Z\bar{Z}}{4}\bdNeq,
\end{equation}
where $\bdNeq = \Braket{(N - \bNeq)^2}{\brhoeq}$ is the variance of the particle number in the dual system.

Therefore, an exponential speedup of the relaxation is obtained by choosing an initial state with no overlap with the charge mode, namely $0 = c(0) = \Braket{c'}{\rho(0)} = \Braket{N}{\rho(0)} -\Neq$, meaning that the initial state should have on average the same number of particle as the equilibrium state.
In that case, Eq.~\eqref{Deq(t)} becomes
\begin{align}\label{Deq(t) only p}
    \Deq^{c=0}[\rho(t)] &= \sum_{k> 1} A_k  \sum_{j=0}^{k}B_{k,k}p(0)^k\e^{-k\gamp t}\\\nonumber
    &\underset{\mathclap{\gamp t \to \infty}}{\sim} \;\;\;\frac{Z\bar{Z}}{2}  p(0)^2 \e^{-2\gamp t},
\end{align}
since $B_{2,2} = Z\bar{Z}$.

\subsection{Mpemba effect}

Let us consider two initial states $\Ket{\varrho_\ii}$ and $\Ket{\varrho_\ii'}$ such that $\Deq(\varrho_\ii') > \Deq(\varrho_\ii)$. Then, at time $t$, the difference in relative entropy of the two states is given by, using Eq.~\eqref{Deq(t)},
\begin{equation}\label{DF21}
    \begin{split}
        &\Deq[\varrho'(t)] - \Deq[\varrho(t)]\\
        &= \sum_{k> 1} A_k  \sum_{j=0}^{k}B_{k,j}(p_\ii'^j c_{\ii}'^{k-j} - p_\ii^j c_{\ii}^{k-j})\e^{-\gamma_{k,j}t},
    \end{split}
\end{equation}
where $c_{\ii} = \Braket{c'}{\varrho_\ii}$, $p_\ii = \Braket{p'}{\varrho_\ii}$, and $c_{\ii}' = \Braket{c'}{\varrho_\ii'}$, $ p_\ii' = \Braket{p'}{\varrho_\ii'}$.

Like before, the slowest decay rate is $\gamma_{2,0}$ and, therefore, if $c_{\ii}'^2  < c_{\ii}^2$, then $\Deq[\varrho'(t)]$ becomes smaller than $\Deq[\varrho(t)]$ in the long time limit. This means that the Mpemba effect occurs, leading to condition \eqref{Mpemba condition}. Furthermore, sufficient conditions to obtain the strong Mpemba effect are $\Deq(\varrho_\ii') > \Deq(\varrho_\ii)$, $c_{\ii}' = 0$ and $c_{\ii}\neq 0$.

In the case where $\abs{c_{\ii}'}  = \abs{c_{\ii}}$, which coefficients we need to look at depends on which of the $\gamma_{k,j}$ is the next slowest rate. Furthermore, this next-slowest decay rate strongly depends on the parameter regime (namely on the choice of $\epsilon, U, T$). Furthermore, the $B_{k,j}$ [Eq.~\eqref{B}] coefficients become more complicated such that it is harder to determine their signs. Note however that $B_{2,1} = 0$, which can be seen by using Eqs.~\eqref{charge} and \eqref{bP_P}.

\section{Additional methods for preparing the initial state}\label{app:prep}
In this appendix, we discuss additional ways of preparing the initial state ${\rho(0)}$ of the dot in complement to the gate-voltage and temperature quenches discussed in Sec.~\ref{sec:init state}. First, we show which initial states are accessible by changing the gate voltage and temperature simultaneously. Then, we consider an interaction-energy quench and, finally, the preparation of an arbitrary nonequilibrium steady state with two reservoirs.

\subsection{Simultaneous gate-voltage switch and temperature quench}\label{app:simultaneous}

\begin{figure}[tb]
    \includegraphics[width=\linewidth]{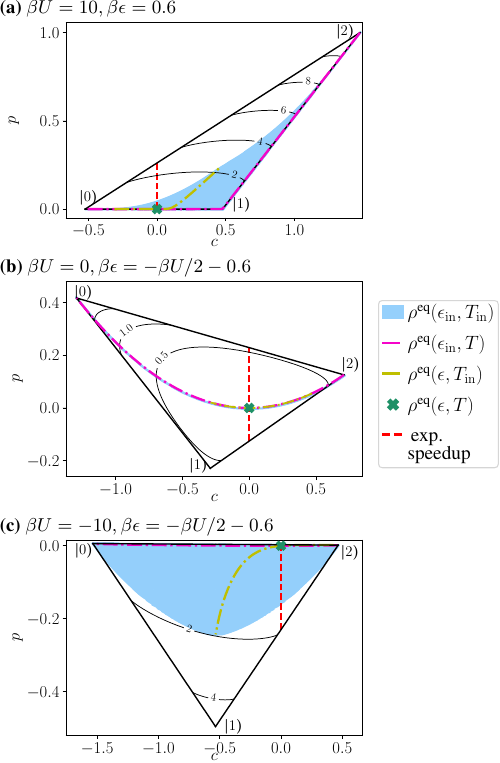}
    \caption{\label{fig:init_eps_T}
        Accessible initial states by simultaneous gate-voltage and temperature quenches. In the triangle of all physical states in the $(c,p)$-plane, the blue area shows the states $\rho(0) = \rho^\eq(\epsilon_\ii, T_\ii)$ that can be prepared by simultaneously quenching the gate voltage and the temperature. We have used the same parameters in the panels (a), (b), (c) as in Figs.~\subfigref{fig:Mpemba}{a-c} and the black lines are the isolines of $\Deq(\rho)$ to make the comparison between the figures easier.
    }
\end{figure}

Following the discussion in Sec.~\ref{sec:combined}, one can wonder to what extent we can widen the range of accessible initial states by applying simultaneously a gate-voltage switch and a temperature quench. In that case, the dot would be initially at equilibrium for energy $\epsilon_\ii$ and temperature $T_\ii$, such that $\rho(0) = \rho^\eq(\epsilon_\ii, T_\ii)$, and then relax toward the new equilibrium state $\rho^\eq(\epsilon, T)$ [Eq.~\eqref{rho_eq}]. We show in Fig.~\ref{fig:init_eps_T} all the initial states (blue area) that can be prepared with this protocol in the same cases as in Figs.~\subfigref{fig:Mpemba}{a-c}. For comparison, we have also indicated the states that can be prepared by a gate-voltage switch alone (magenta dash-dotted line), as described in Sec.~\ref{sec:energy quench}, or by a temperature quench alone (yellow dash-dotted line), as described in Sec.~\ref{sec:temperature quench}. As expected, for strong interaction, in Figs.~\subfigref{fig:init_eps_T}{a} and \subfigref{fig:init_eps_T}{c}, this combination of quenches gives access to more initial states, including states that will relax with an exponential speedup (dashed red line) since they have a vanishing charge amplitude, $c(0) = 0$. On the contrary, in the noninteracting case, in Fig.~\subfigref{fig:init_eps_T}{b}, there is no benefit in combining the quenches as all the prepared initial states fall on the same line and could already be obtained by a gate switch alone.

\begin{figure*}[!]
    \includegraphics[width=0.99\linewidth]{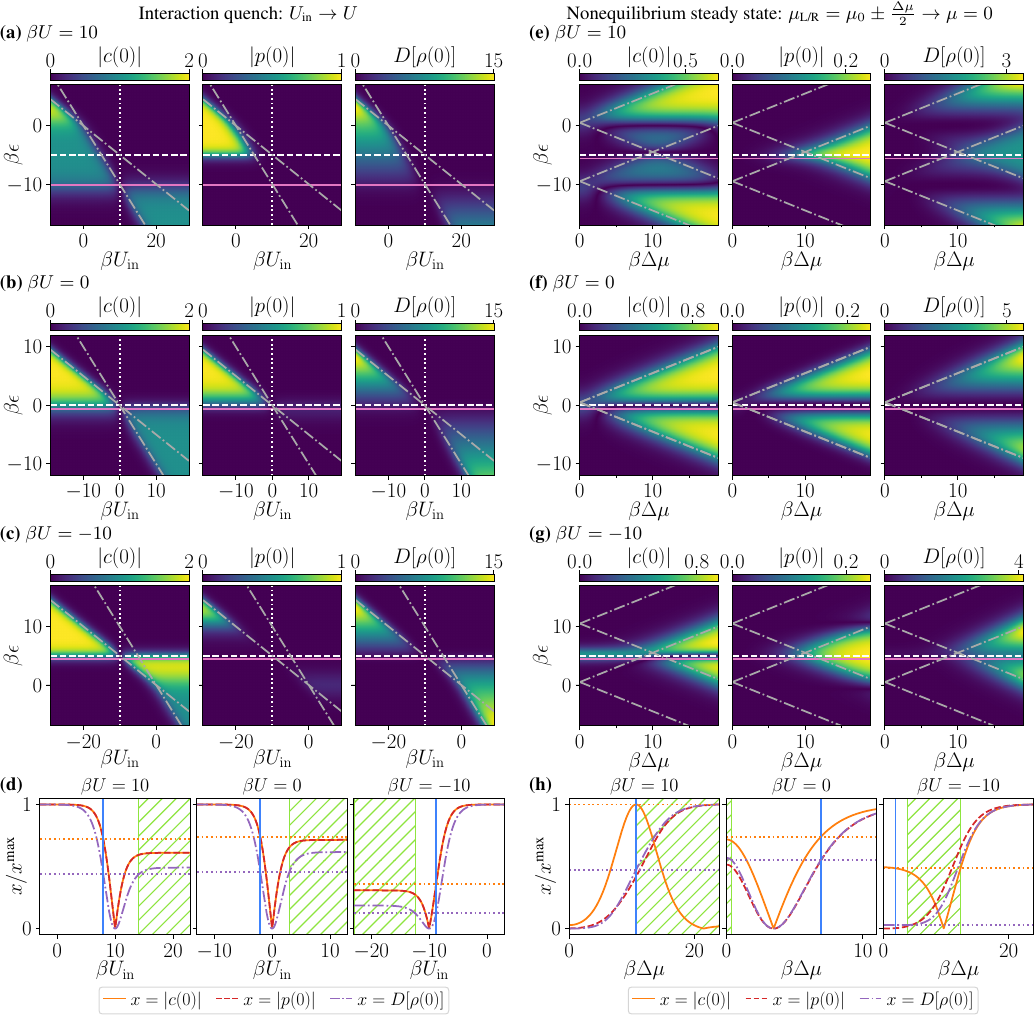}\vspace{-0.2cm}
    \caption{\label{fig:U quench}\label{fig:DV}
        \textit{Left column:} State preparation with an interaction-energy quench.
        The absolute values of the charge amplitude and parity amplitude, as well as the relative entropy (with respect to the equilibrium state at energy $\epsilon$ and temperature $T$) of the prepared initial state $\rho(0) = \rhoeq(U_\ii)$ are plotted as functions of $U_\ii$ and $\epsilon$ for \textbf{(a)} repulsive interaction, \textbf{(b)} no interaction and \textbf{(c)} attractive interaction. The dotted white line corresponds to $\rho(0) = \rhoeq(U)$ and the dashed white line indicates the electron-hole symmetric point $\epsilon = -U/2$. The gray dash-dotted lines provide a guide to the eye and correspond to $\epsilon = -U_\ii/2$ and $\epsilon = -U_\ii$.
        \textbf{(d)} Slices of (a), (b) and (c) at the energy indicated by the solid pink lines,  $\beta\epsilon = - \beta U - 0.2$ for $U > 0$ and $\beta\epsilon = -\beta U/2  - 0.6 $ for $U \le 0$. All quantities are normalized by their maximum. The vertical blue line indicates the interaction value $\mathcal{U}_\ii$ used to prepare $\varrho_\ii$ and the area hashed in green corresponds to the range of values of $\mathcal{U}_{\ii}'$ that can be used to prepare an initial state $\varrho'_\ii$ that will give a Mpemba effect compared to the initial state $\varrho_\ii$.
        \textit{Right column:} State preparation with an initial nonequilibrium steady-state.
        The absolute values of the charge amplitude and parity amplitude, as well as the relative entropy (with respect to the equilibrium state at energy $\epsilon$ and temperature $T$) of the prepared initial state $\rho(0) = \rho^\text{ss}(\Delta\mu)$ are plotted as functions of $\Delta \mu$ and $\epsilon$ for $\beta \mu_0 = 0.5$ and \textbf{(e)} repulsive interaction, \textbf{(f)} no interaction and \textbf{(g)} attractive interaction. The dashed white line indicates the electron-hole symmetric point $\epsilon = -U/2$. The gray dash-dotted lines indicate the resonances $\epsilon = \mu_\text{L/R}$ and $\epsilon + U = \mu_\text{L/R}$. \textbf{(h)} Slices of (e), (f) and (g) at the energy indicated by the solid pink lines, $\beta\epsilon = -\beta U/2  - 0.6$. The vertical blue line indicates the bias $\Delta \mu_\ii$ used to prepare $\varrho_\ii$ and the area hashed in green corresponds to the range of values of $\Delta \mu_\ii'$ that can be used to prepare an initial state $\varrho'_\ii$ that will give a Mpemba effect compared to the initial state $\varrho_\ii$. In panels (d) and (h), the dotted orange and purple lines are guides to the eye indicating respectively $\abs{c_{\ii}}$ and $\Deq(\varrho_\ii)$, that is the values of $\abs{c(0)}$ and $D[\rho(0)]$ at the vertical blue line so we can see that the area in green fulfills the Mpemba conditions \eqref{Mpemba condition}.
    }
\end{figure*}

\subsection{Interaction quench}

For completeness, we consider here an interaction-energy quench. The dot is initially at equilibrium for interaction energy $U_\ii$, then at time $t=0$ the interaction energy is instantaneously set to a new value $U$.
This means that the prepared initial state is $\rho(0) = \rhoeq(U_\ii) = \exp(-\beta H_{U_\ii})/Z_{U_\ii}$, where $H_{U_\ii}$ is the dot Hamiltonian as given in Eq.~\eqref{H_dot}, but with interaction energy $U_\ii$ and $Z_{U_\ii}$ the corresponding partition function.
We plot in Figs.~\subfigref{fig:U quench}{a-c} the charge amplitude, parity amplitude and relative entropy of this initial state as functions of $U_\ii$ and $\epsilon$. We notice in all the plots, for all kinds of interaction, features with a slope $\epsilon = -U_\ii/2$ for attractive initial interactions ($U_\ii < 0$) which changes to a slope $\epsilon = -U_\ii$ when the initial interaction becomes repulsive (the gray dash-dotted lines provide a guide to the eye). This is because the two-particle resonance ($\epsilon + U_\ii/2 = \mu$, with here $\mu = 0$) occurring for attractive interaction is replaced by a resonance of the transition energy $\epsilon + U_\ii$ with the reservoir when the interaction becomes repulsive~\cite{Monsel2022Jul}.

First, by comparing the plots for the charge and parity amplitude, we can see that an interaction-energy quench does not allow for preparing an initial state giving an exponential speedup.
Second, looking at the slices plotted in Fig.~\subfigref{fig:U quench}{d}, we see that for repulsive interaction, $\beta U \gg 1$, the Mpemba effect can be realized by preparing $\varrho_\ii = \rhoeq(\mathcal{U}_\ii)$, where we have chosen $\mathcal{U}_{\ii} < U$, and by then finding a value $\mathcal{U}_{\ii}'>U$  to prepare $\varrho_\ii' = \rhoeq(\mathcal{U}_\ii')$. This works exclusively for $\epsilon \simeq -U$ but not for other values of $\epsilon$.  There is in particular a range of values of $\mathcal{U}_\ii$, including the one chosen in Fig.~\subfigref{fig:U quench}{d} (blue vertical line) such that $\Deq(\varrho_\ii) < \lim_{\mathcal{U}_\ii'\to+\infty}\Deq(\varrho_\ii')$ and $\abs{c_{\ii}} > \lim_{\mathcal{U}_\ii'\to+\infty}\abs{c'_{\ii}}$ which allows choosing any $\mathcal{U}_\ii'$ larger than a threshold value of $U_\ii$ fulfilling $U_\ii > U$ and $\Deq[\rhoeq(U_\ii)] = \Deq(\varrho_\ii)$. We find similar results for the weak interaction case.
On the contrary, for large attractive interaction, the Mpemba effect can be realized by taking $\mathcal{U}_\ii < U$ and $\mathcal{U}_\ii'>U$ for $\epsilon$ close to the particle-hole symmetry point but not for values of $\epsilon$ far away from it. It is particularly interesting to choose $\mathcal{U}_\ii$ such that $\Deq(\varrho_\ii) < \lim_{\mathcal{U}_\ii'\to-\infty}\Deq(\varrho_\ii')$ and $\abs{c_{\ii}} > \lim_{\mathcal{U}_\ii'\to-\infty}\abs{c'_{\ii}}$ which allows choosing any $\mathcal{U}_\ii'$ below a threshold value.

\subsection{Nonequilibrium steady state}

We prepare the initial state by coupling, for $t < 0$, two reservoirs, that we will call left and right, with temperatures $T_{\L/\R}$ and electrochemical potentials $\mu_{\L/\R}$, such that $\Ket{\rho(0)}$ is the nonequilibrium steady state $\Ket{\rho^\text{ss}}$ corresponding to this two-terminal setup \cite{Monsel2022Jul}. For simplicity, we investigate either the case of a potential bias $\Delta \mu$ or the case of a temperature bias $\Delta T = T_\L - T_\R$. At $t = 0$, we bring the temperatures of both reservoirs to $T$ and their electrochemical potentials to $\mu = 0$, such that we get a situation equivalent to the one studied in the main text where a single reservoir is coupled to the dot.

First, note that having one of the two reservoirs of the initial-state preparation at the same temperature and chemical potential as the final reservoir, e.g., $T_\R = T$ and $\mu_\R = 0$, is very similar to the situation of having a single reservoir. Therefore changing $\mu_\L$ gives similar results as the gate-voltage switches discussed in Sec.~\ref{sec:energy quench} and changing $T_\L$ gives similar results as the temperature quench from Sec.~\ref{sec:temperature quench}.

We studied the case $\mu_\L = \mu_\R = 0$ but $T_\L = T_\R +\Delta T$. The features of the amplitudes and relative entropy as functions of $\Delta T$ and $\epsilon$ remain qualitatively very similar to Figs.~\subfigref{fig:T quench}{e-g} (single bath temperature quench, see Sec.~\ref{sec:temperature quench}), even when $T_\R$ is not equal to $T$. We therefore do not show explicit results for this case here.

Instead, we focus on the case of a potential bias, $\mu_{\L/\R} = \mu_0 \pm \Delta\mu/2$, but with equal temperatures, $T_\L = T_\R = T$.
In the case where $\mu_0 = 0$, only strong repulsive interaction allows for preparing states to get the Mpemba effect thanks to a local maximum of $\abs{c_\ii}$, like in Sec.~\ref{sec:temperature quench} for the temperature quench. The symmetry of all the curves prevents us from finding initial states fulfilling the Mpemba conditions. Therefore, we will study in more detail the case where the electrochemical potential are shifted by $\mu_0 \neq 0$, such that the setup is not symmetrical around the chemical potential at $t < 0$. This means that at $t=0$, we switch off the bias, $\Delta \mu \to 0$, while also bringing $\mu_0$ to 0.
We have plotted in Figs.~\subfigref{fig:DV}{e-g} the charge amplitude, parity amplitude and relative entropy of the prepared initial state $\rho(0) = \rho^\text{ss}(\Delta\mu)$ as functions of $\Delta \mu$ and $\epsilon$ for $\beta\mu_0 = 0.5$. Note that the plots for $\Delta \mu \le 0$ are the mirror-symmetric of the ones for $\Delta \mu \ge 0$ that we are showing.
We have indicated by dash-dotted gray lines when the dot energy transitions $\epsilon$ and $\epsilon + U$ are resonant with one of the reservoirs, namely equal to $\mu_\L$ or $\mu_\R$.
Looking at the slices at $\beta \epsilon = -\beta U /2 - 0.6$ plotted in Fig.~\subfigref{fig:DV}{h}, the most favorable case is the strong repulsive interaction, where $\varrho_\ii = \rho^\text{ss}(\Delta\mu_\ii)$ can be prepared with the potential bias corresponding to the maximum of $\abs{c}$.
Then, any $\Delta\mu_\ii' > \Delta\mu_\ii$ gives a state $\varrho_\ii' = \rho^\text{ss}(\Delta\mu_\ii')$ fulfilling the Mpemba condition \eqref{Mpemba condition}. For strong attractive interaction, only a restricted range of values of $\Delta\mu_\ii'$ fulfill the Mpemba condition and the least favorable case is the weak interaction, as shown by the narrowness of the green-hashed area corresponding to possible values of $\Delta \mu_\ii'$ for the chosen $\Delta\mu_\ii$ (vertical blue line) in the central plot of Fig.~\subfigref{fig:DV}{h}. Finally, an interesting feature for strong interactions, $\beta\abs{U} \gg 1$, is that $|c|$ goes to 0 at a bias $\Delta \mu_\ii''$ where the parity amplitude is finite. This point would therefore allow the preparation of an initial state $\varrho_\ii'' = \rho^\text{ss}(\Delta\mu_\ii'')$ giving the strong Mpemba effect with respect to $\varrho_\ii$.

\bibliography{Refs.bib}

\end{document}